\documentclass{article}
\usepackage{amssymb}
\usepackage{amsfonts}
\usepackage{amsmath}

\usepackage{eurosym}
\usepackage{amssymb}
\usepackage{amsfonts}
\usepackage{amsmath}
\usepackage{graphicx}
\usepackage[ruled]{algorithm2e}
\usepackage{subcaption}
\usepackage{xcolor}

\usepackage{titlesec}
\usepackage[titletoc,toc,title]{appendix}

\begin{document}

\title{Parallel Concatenation of Bayesian Filters: Turbo Filtering}
\author{}
\maketitle

\begin{abstract}
In this manuscript a method for developing novel filtering algorithms
through the parallel concatenation of two Bayesian filters is illustrated.
Our description of this method, called turbo filtering, is based on a new
graphical model; this allows us to efficiently describe both the processing
accomplished inside each of the constituent filter and the interactions between
them. This model is exploited to develop two new filtering algorithms for
conditionally linear Gaussian systems. Numerical results for a specific
dynamic system evidence that such filters can achieve a better
complexity-accuracy tradeoff than marginalized particle filtering.
\end{abstract}

\bigskip

\begin{center}
Giorgio M. Vitetta, Pasquale Di Viesti, Emilio Sirignano and Francesco Montorsi

\vspace{5mm}University of Modena and Reggio Emilia

Department of Engineering "Enzo Ferrari"

Via P. Vivarelli 10/1, 41125 Modena - Italy

email: giorgio.vitetta@unimore.it, 190010@studenti.unimore.it,
emilio.sirignano@unimore.it, francesco.montorsi@gmail.com
\end{center}

\bigskip


\bigskip\vspace{1cm}

\baselineskip0.2 in\newpage

\section{Introduction\label{sec:intro}}

The \emph{nonlinear filtering problem} consists of inferring the \emph{%
posterior distribution} of the hidden state of a nonlinear dynamic system
from a set of past and present measurements \cite{Arulampalam_2002}. A
general recursive solution to this problem, known as \emph{Bayesian filters%
} (e.g., see \cite[Sect. II, eqs. (3)-(5)]{Arulampalam_2002}), is available,
but, unluckily, can be put in closed form in few cases \cite{Anderson_1979}.
In the past, various filtering methods generating a \emph{functional
approximation} of the desired posterior pdf have been developed; these can
be divided into \emph{local} and \emph{global} methods on the basis of the
way the posterior pdf is approximated \cite{Mazuelas_2013}, \cite{Smidl_2008}%
. On the one hand, local techniques, like \emph{extended Kalman filtering}
(EKF) \cite{Anderson_1979}, are computationally efficient, but may suffer
from error accumulation over time; on the other hand, global techniques,
like \emph{sequential Monte Carlo }(SMC) algorithms \cite{Doucet_2001}, \cite%
{Doucet_2000} (also known as \emph{\ particle filtering}, PF \cite%
{Gustafsson_et_al_2002}, \cite{Andrieu_2002}) may achieve high accuracy at
the price, however, of unacceptable complexity and numerical problems. These
considerations have motivated the investigation of other methods able to
achieve high accuracy under given computational constraints. Some of such
solutions are based on the idea of \emph{combining} (i.e., \emph{%
concatenating}) \emph{local and global methods}; relevant examples of this
approach are represented by a) \emph{marginalized particle filtering} (MPF) 
\cite{Schon_2005} and other techniques related to it (e.g., see \cite%
{Smidl_2008} and \cite{Lu_2007}) and b) \emph{cascaded architectures} based
on the joint use of EKF and PF (e.g., see \cite{Krach_2008_bis} and \cite%
{Vitetta_2014}). Note that, in all these cases, two heterogeneous methods
are combined in a way that the resulting filtering algorithm is \emph{%
forward only} and,\ within its recursion, each of such methods is executed
only once;\ for this reason, if the jargon of \emph{coding theory} is
adopted in this context, such filtering algorithms can be seen as specific
instances of the general concept of \emph{serial concatenation} \cite%
{Vitetta}, \cite{Benedetto_1998} of two (\emph{constituent}) filtering
methods.

In this manuscript, we focus on the novel concept of \emph{parallel
concatenation} (PC) of Bayesian filterings, i.e. on the idea of combining
two (constituent) filters in a way that, within each recursion of the
resulting concatenated algorithm, they can iteratively refine their
statistical information through the mutual exchange of probabilistic (i.e., 
\emph{soft}) information; this concept is dubbed \emph{turbo filtering} (TF)
for its resemblance to the iterative (i.e., \emph{turbo}) decoding of
concatenated channel codes \cite{Berrou_1996}. More specifically, we first
develop a \emph{general graphical model} that allows us to: a) represent the
PC of two Bayesian filters as the interconnection of \emph{two soft-in
soft-out }(SISO) modules, b) represent the iterative processing accomplished
by these modules as a message passing technique and c) to derive the
expressions of the passed messages by applying the \emph{sum-product
algorithm} (SPA) \cite{Loeliger_2007}, \cite{Kschischang_2001}, together
with a specific \emph{scheduling} procedure, to the graphical model itself.
Then, the usefulness of this approach is exemplified by developing two TF\
algorithms for the class of \emph{conditionally linear Gaussian} (CLG) SSMs 
\cite{Schon_2005}. Our computer simulations for a specific CLG\ SSM evidence
that, in the considered case, these algorithms perform very closely to MPF,
but are substantially faster.

It is worth mentioning that the TF principle has been formulated for the
first time in \cite{Montorsi_2013}, where it has also been successfully
applied to inertial navigation. However, all the theoretical results
illustrated in this manuscript have been obtained later and have been
inspired by various results available in the literature about: a) the
representation of filtering methods as \emph{message passing procedures on
factor graphs} (e.g., see \cite{Loeliger_2007}, \cite{Kschischang_2001} and 
\cite{Dauwels_2006}); b) the use of \ graphical models in the derivation and
interpretation of \emph{turbo decoding} and \emph{turbo equalization} \cite%
{Loeliger_2007}, \cite{Kschischang_2001}, \cite{Worthen_2001}.

The remaining part of this manuscript is organized as follows. A description
of the considered SSM is illustrated in Section \ref{sec:scenario}. In
Section \ref{sec:Factorgraphs} a new graphical model describing the TF\
principle is devised; then, a specific case of that model, referring to the
use of an extended Kalman filter and particle filter as constituent filters,
and a CLG SSM is analysed. The derivation of two TF algorithms based on the
last model is illustrated in Section \ref{sec:Message-Passing}, whereas
their interpretation from a coding theory perspective is discussed in
Section \ref{Interpretation}. Such algorithms are compared with EKF and MPF,
in terms of accuracy and execution time, in Section \ref{num_results}.
Finally, some conclusions are offered in Section \ref{sec:conc}.

\section{Model Description\label{sec:scenario}}

In the following we focus on a discrete-time CLG SSM \cite{Schon_2005},
whose $D$-dimensional \emph{hidden state} $\mathbf{x}_{l}\triangleq \lbrack
x_{0,l},x_{1,l},...,$ $x_{D-1,l}]^{T}$ in the $l$-th interval is partitioned
as $\mathbf{x}_{l}=[(\mathbf{x}_{l}^{(L)})^{T},(\mathbf{x}%
_{l}^{(N)})^{T}]^{T}$; here, $\mathbf{x}_{l}^{(L)}\triangleq \lbrack
x_{0,l}^{(L)}$, $x_{1,l}^{(L)},...,x_{D_{L}-1,l}^{(L)}]^{T}$ ($\mathbf{x}%
_{l}^{(N)}\triangleq \lbrack
x_{0,l}^{(N)},x_{1,l}^{(N)},$ $...,x_{D_{N}-1,l}^{(N)}]^{T}$) is the so called 
\emph{linear }(\emph{nonlinear}) \emph{component} of $\mathbf{x}_{l}$, with $%
D_{L}<D$ ($D_{N}=D-D_{L}$). Following \cite{Schon_2005} and \cite{Lu_2007},
the models 
\begin{equation}
\mathbf{x}_{l+1}^{(Z)}=\mathbf{A}_{l}^{(Z)}\left( \mathbf{x}%
_{l}^{(N)}\right) \mathbf{x}_{l}^{(L)}+\mathbf{f}_{l}^{(Z)}\left( \mathbf{x}%
_{l}^{(N)}\right) +\mathbf{w}_{l}^{(Z)}  \label{eq:XL_update}
\end{equation}%
and%
\begin{eqnarray}
\mathbf{y}_{l} &\triangleq &[y_{0,l},y_{1,l},...,y_{P-1,l}]^{T}  \notag \\
&=&\mathbf{g}_{l}\left( \mathbf{x}_{l}^{(N)}\right) +\mathbf{B}_{l}\left( 
\mathbf{x}_{l}^{(N)}\right) \mathbf{x}_{l}^{(L)}+\mathbf{e}_{l}
\label{eq:y_t}
\end{eqnarray}%
are adopted for the update of the \emph{linear} ($Z=L$) and \emph{nonlinear}
($Z=N$) components, and for the $P$-dimensional vector of noisy measurements
available in the $l$-th interval, respectively. In the \emph{state update
model} (\ref{eq:XL_update}) $\mathbf{f}_{l}^{(Z)}\left( \mathbf{x}\right) $ (%
$\mathbf{A}_{l}^{(Z)}(\mathbf{x}_{l}^{(N)})$) is a time-varying $D_{Z}$%
-dimensional real function ($D_{Z}\times D_{L}$ real matrix) and $\mathbf{w}%
_{l}^{(Z)}$ is the $l$-th element of the process noise sequence $\{\mathbf{w}%
_{k}^{(Z)}\}$; this sequence consists of $D_{Z}$-dimensional \emph{%
independent and identically distributed} (iid) Gaussian noise\emph{\ }%
vectors, each characterized by a zero mean and a covariance matrix $\mathbf{C%
}_{w}^{(Z)}$ (independence between $\{\mathbf{w}_{k}^{(L)}\}$ and $\{\mathbf{%
w}_{k}^{(N)}\}$ is also assumed for simplicity). Moreover, in the \emph{%
measurement model} (\ref{eq:y_t}), $\mathbf{B}_{l}(\mathbf{x}_{l}^{(N)})$ is
a time-varying $P\times D_{L}$ real matrix, $\mathbf{g}_{l}(\mathbf{x}%
_{l}^{(N)})$ is a time-varying $P$-dimensional real function and $\mathbf{e}%
_{l}$ the $l$-th element of the measurement noise sequence $\left\{ \mathbf{e%
}_{k}\right\} $; this sequence consists of $P$-dimensional iid Gaussian
noise vectors (each characterized by a zero mean and a covariance matrix $%
\mathbf{C}_{e}$), and is independent of both $\{\mathbf{w}_{k}^{(N)}\}$ and $%
\{\mathbf{w}_{k}^{(L)}\}$.

In the following we take into consideration not only the detailed models (%
\ref{eq:XL_update}) and (\ref{eq:y_t}), but also their more compact
counterparts%
\begin{equation}
\mathbf{x}_{l+1}=\mathbf{f}_{l}\left( \mathbf{x}_{l}\right) +\mathbf{w}_{l}
\label{eq:X_update}
\end{equation}%
and%
\begin{equation}
\mathbf{y}_{l}=\mathbf{h}_{l}\left( \mathbf{x}_{l}\right) +\mathbf{e}_{l}
\label{meas_mod}
\end{equation}%
respectively, which refer to the whole state; here, $\mathbf{f}_{l}\left( 
\mathbf{x}_{l}\right) $ ($\mathbf{w}_{l}$) is a $D$-dimensional function
(Gaussian noise vector\footnote{%
The covariance matrix $\mathbf{C}_{w}$ of $\mathbf{w}_{l}$ can be easily
computed on the basis of the matrices $\mathbf{C}_{w}^{(L)}$ and $\mathbf{C}%
_{w}^{(N)}$.}) deriving from the ordered concatenation of the vectors $%
\mathbf{A}_{l}^{(L)}(\mathbf{x}_{l}^{(N)})\mathbf{x}_{l}^{(L)}+\mathbf{f}%
_{l}^{(L)}(\mathbf{x}_{l}^{(N)})$ and $\mathbf{A}_{l}^{(N)}(\mathbf{x}%
_{l}^{(N)})\mathbf{x}_{l}^{(L)}+\mathbf{f}_{l}^{(N)}(\mathbf{x}_{l}^{(N)})$ (%
$\mathbf{w}_{l}^{(L)}$ and $\mathbf{w}_{l}^{(N)}$; see (\ref{eq:XL_update}%
)), and $\mathbf{h}_{l}\left( \mathbf{x}_{l}\right) \triangleq \mathbf{g}%
_{l}(\mathbf{x}_{l}^{(N)})+\mathbf{B}_{l}(\mathbf{x}_{l}^{(N)})\mathbf{x}%
_{l}^{(L)}$. Moreover, since EKF is employed in the TF algorithms developed
in the following, the \emph{linearized} versions of (\ref{eq:X_update}) and (%
\ref{meas_mod}) are also considered; these can be expressed as (e.g., see 
\cite[pp. 194-195]{Anderson_1979})%
\begin{equation}
\mathbf{x}_{l+1}=\mathbf{F}_{l}\mathbf{x}_{l}+\mathbf{u}_{l}+\mathbf{w}_{l}
\label{state_up_approx}
\end{equation}%
and%
\begin{equation}
\mathbf{y}_{l}=\mathbf{H}_{l}^{T}\mathbf{x}_{l}+\mathbf{v}_{l}+\mathbf{e}%
_{l},  \label{meas_mod_approx}
\end{equation}%
respectively; here, $\mathbf{F}_{l}\triangleq \lbrack \partial \mathbf{f}%
_{l}\left( \mathbf{x}\right) /\partial \mathbf{x}]_{\mathbf{x=x}_{fe,l}}$, $%
\mathbf{x}_{fe,l}$ is the (forward) estimate of $\mathbf{x}_{l}$ evaluated
by EKF in its $l$-th recursion, $\mathbf{u}_{l}\triangleq \mathbf{f}%
_{l}\left( \mathbf{x}_{fe,l}\right) -\mathbf{F}_{l}\mathbf{x}_{fe,l}$, $%
\mathbf{H}_{l}^{T}\triangleq \lbrack \partial \mathbf{h}_{l}\left( \mathbf{x}%
\right) /\partial \mathbf{x}]_{\mathbf{x=x}_{fp,l}}$, $\mathbf{x}_{fp,l}$ is
the (forward) prediction $\mathbf{x}_{l}$ computed by EKF in its $(l-1)$-th
recursion and $\mathbf{v}_{l}\triangleq \mathbf{h}_{l}\left( \mathbf{x}%
_{fp,l}\right) -\mathbf{H}_{l}^{T}\mathbf{x}_{fp,l}$.

In the following Section we focus on the so-called \emph{filtering problem},
which concerns the evaluation of the posterior pdf $f(\mathbf{x}_{l}|\mathbf{%
y}_{1:t})$ at an instant $t\geq 1$, given a) the initial pdf $f(\mathbf{x}%
_{1})$ and b) the $t\cdot P$-dimensional \emph{measurement} vector $\mathbf{y%
}_{1:t}=\left[ \mathbf{y}_{1}^{T},\mathbf{y}_{2}^{T},...,\mathbf{y}_{t}^{T}%
\right] ^{T}$.

\section{Graphical Modelling for Turbo Filtering\label{sec:Factorgraphs}}

\begin{figure}[tbp]
\centering
\includegraphics[width=0.80
\textwidth]{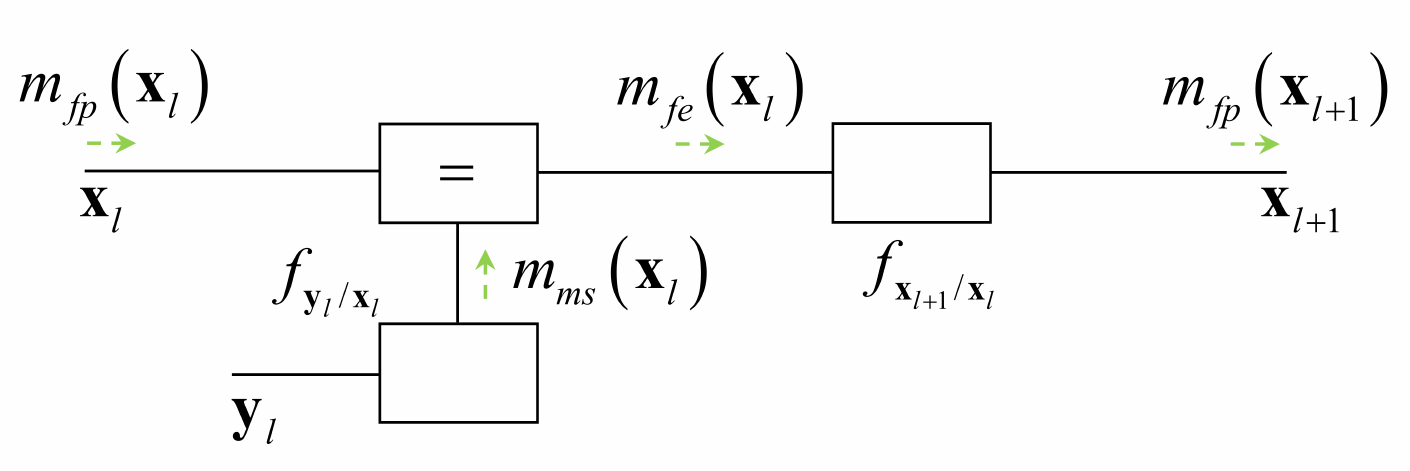}
\caption{Factor graph representing the $l$-th recursion of Bayesian
filtering for a SSM described by the Markov model $f(\mathbf{x}_{l+1}|%
\mathbf{x}_{l})$ and the observation model $f(\mathbf{y}_{l}|\mathbf{x}_{l})$%
; the SPA message flow is indicated by green arrows.}
\label{Fig_1}
\end{figure}

Let us consider first a SSM described by the \emph{Markov model} $f(\mathbf{x%
}_{l+1}|\mathbf{x}_{l})$ and the \emph{observation model} $f(\mathbf{y}_{l}|%
\mathbf{x}_{l})$ for any $l$. In this case, the computation of the posterior
pdf $f(\mathbf{x}_{t}|\mathbf{y}_{1:t})$ for $t\geq 1$ can be accomplished
by means of an exact \emph{Bayesian recursive procedure}, consisting of a 
\emph{measurement update} (MU) step followed by a \emph{time update} (TU)
step. Following \cite[Sec. II, p. 1297]{Loeliger_2007}, the equations
describing the $l$-th recursion of this procedure (with $l=1,2,...,t$) can
be\ easily obtained by applying the SPA to the Forney-style\emph{\ }FG shown
in Fig. \ref{Fig_1}, if the joint pdf $f(\mathbf{x}_{t},\mathbf{y}_{1:t})$
is considered in place of the associated a posteriori pdf $f(\mathbf{x}_{t}|%
\mathbf{y}_{1:t})$. In fact, given the measurement message $\vec{m}%
_{ms}\left( \mathbf{x}_{l}\right) =f\left( \mathbf{y}_{l}\left\vert \mathbf{x%
}_{l}\right. \right) $, if the input message\footnote{%
In the following the acronyms \emph{fp} and \emph{fe} are employed in the
subscripts of various messages, so that readers can easily understand their
meaning; in fact, the messages these acronyms refer to represent a form of
one-step \emph{forward prediction} and of \emph{forward estimation},
respectively.} $\vec{m}_{fp}\left( \mathbf{x}_{l}\right) =f(\mathbf{x}_{l},%
\mathbf{y}_{1:(l-1)})$ enters this FG, the message going out of the \emph{%
equality node} is given by%
\begin{eqnarray}
\vec{m}_{fe}\left( \mathbf{x}_{l}\right) &=&\vec{m}_{fp}\left( \mathbf{x}%
_{l}\right) \vec{m}_{ms}\left( \mathbf{x}_{l}\right)  \notag \\
&=&f(\mathbf{x}_{l},\mathbf{y}_{1:(l-1)})f\left( \mathbf{y}_{l}\left\vert 
\mathbf{x}_{l}\right. \right) =f(\mathbf{x}_{l},\mathbf{y}_{1:l})
\label{meas_update}
\end{eqnarray}%
and, consequently, the message emerging from the \emph{function node}
referring to the pdf $f(\mathbf{x}_{l+1}|\mathbf{x}_{l})$ is expressed by%
\begin{equation}
\int f\left( \mathbf{x}_{l+1}\left\vert \mathbf{x}_{l}\right. \right) \vec{m}%
_{fe}\left( \mathbf{x}_{l}\right) d\mathbf{x}_{l}=f(\mathbf{x}_{l+1},\mathbf{%
y}_{1:l})=\vec{m}_{fp}\left( \mathbf{x}_{l+1}\right) .  \label{time_update}
\end{equation}%
Eqs. (\ref{meas_update}) and (\ref{time_update}) express the MU and the TU,
respectively, that need to be accomplished in the $l$-th recursion of
Bayesian filtering.

\begin{figure}[tbp]
\centering
\includegraphics[width=0.80%
\textwidth]{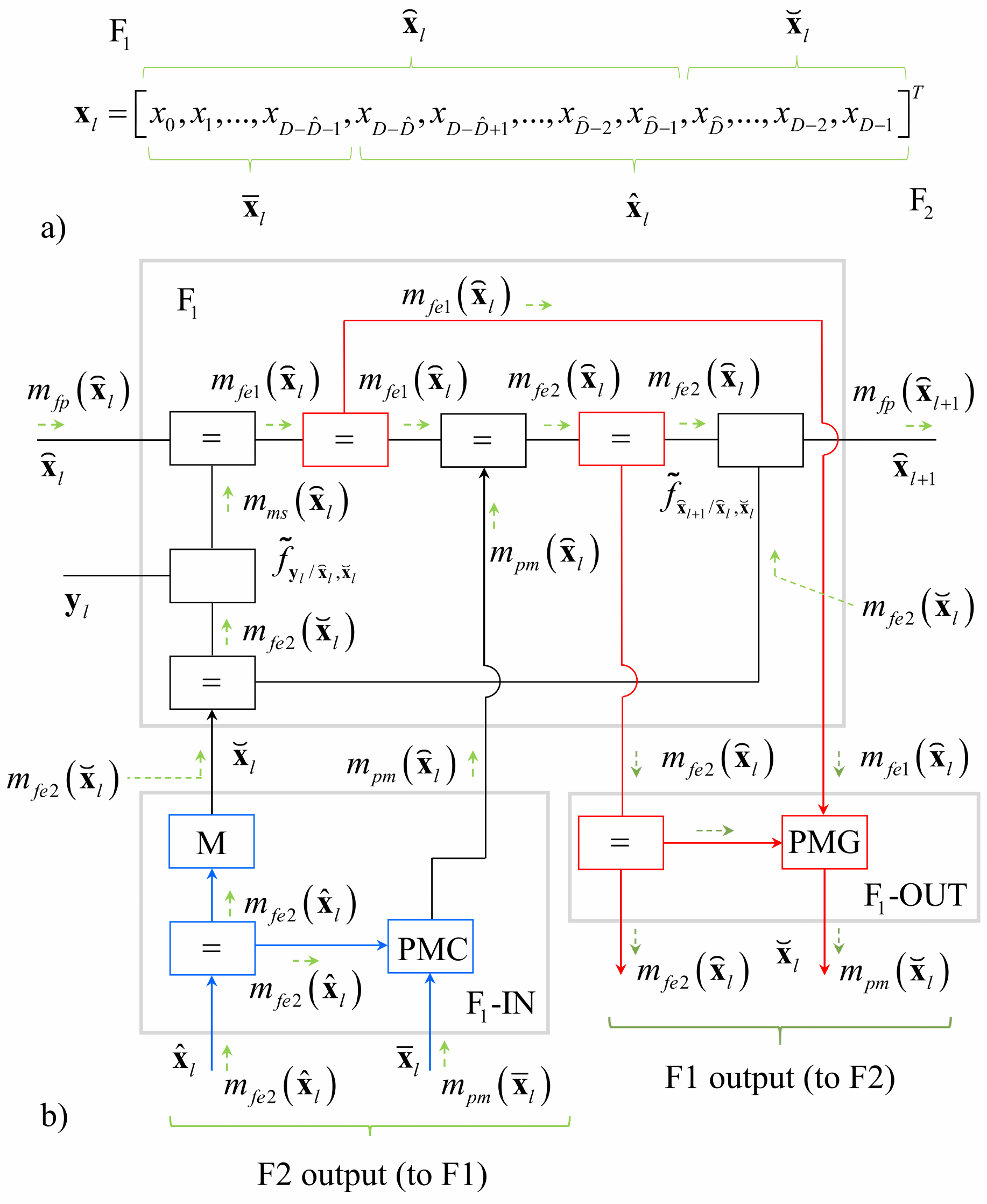}
\caption{a) Partitioning adopted for the system state $\mathbf{x}_{l}$ in
the PC of two filtering algorithms; b) Graphical model referring to a SISO
module based on F$_{1}$. Black and blue (red) lines are used to identify the
edges and the blocks related to filtering and processing of information
coming from F$_{2}$ (to be delivered to F$_{2}$), respectively.}
\label{Fig_2}
\end{figure}

Let us see now how the FG illustrated in Fig. \ref{Fig_1} can be exploited
to devise a graphical model efficiently representing the TF concept. As
already stated in the Introduction, any TF scheme results from the \emph{%
parallel concatenation }of two \emph{constituent} Bayesian filters (denoted F%
$_{1}$ and F$_{2}$ in the following), that can iteratively improve their
accuracy through the exchange of their statistical information. In practice,
in developing TF techniques, the following general rules are followed: 
\textbf{R1}) the constituent filters operate on \emph{partially overlapped}
portions of system state; \textbf{R2}) the filter F$_{1}$ (F$_{2}$) is the
core of a \emph{processing module} (called \emph{soft-in soft-out}, SISO,
module in the following) receiving statistical information from F$_{2}$ (F$%
_{1}$) and generating new statistical information useful to F$_{2}$ (F$_{1}$%
); \textbf{R3}) each constituent filter relies on \emph{exact}
Markov/observation models or \emph{approximate} (e.g., linearized) versions
of them. These rules can be motivated and implemented as follows. The \emph{%
first rule} (i.e., \textbf{R1}) ensures that any TF filtering algorithm
contains a form of \emph{redundancy}, that represents the first of the two
fundamental properties characterizing each \emph{error correction method}
employed in digital communications \cite{Vitetta}. In our general
description of a TF scheme, it is assumed that (see Fig. \ref{Fig_2}-(a)):
1) filter F$_{1}$ (F$_{2}$) estimates the state vector $\overset{%
\mathchoice{}{}{\mbox{\raisebox{-.5ex}[0pt][0pt]{$\scriptstyle
\smallfrown$}}}{\mbox{\raisebox{-.35ex}[0pt][0pt]{$\scriptscriptstyle
\smallfrown$}}}}{\mathbf{x}}_{l}$ ($\mathbf{\hat{x}}_{l}$) of size $\overset{%
\mathchoice{}{}{\mbox{\raisebox{-.5ex}[0pt][0pt]{$\scriptstyle
\smallfrown$}}}{\mbox{\raisebox{-.35ex}[0pt][0pt]{$\scriptscriptstyle
\smallfrown$}}}}{D}$ ($\hat{D}$), with $\overset{%
\mathchoice{}{}{\mbox{\raisebox{-.5ex}[0pt][0pt]{$\scriptstyle
\smallfrown$}}}{\mbox{\raisebox{-.35ex}[0pt][0pt]{$\scriptscriptstyle
\smallfrown$}}}}{D}\leq D$ ($\hat{D}\leq D$); 2) the portion $\overset{%
\mathchoice{}{}{\mbox{\raisebox{-.5ex}[0pt][0pt]{$\scriptstyle
\smallsmile$}}}{\mbox{\raisebox{-.35ex}[0pt][0pt]{$\scriptscriptstyle
\smallsmile$}}}}{\mathbf{x}}_{l}$ ($\mathbf{\bar{x}}_{l}$) of $\mathbf{x}%
_{l} $ not included in $\overset{%
\mathchoice{}{}{\mbox{\raisebox{-.5ex}[0pt][0pt]{$\scriptstyle
\smallfrown$}}}{\mbox{\raisebox{-.35ex}[0pt][0pt]{$\scriptscriptstyle
\smallfrown$}}}}{\mathbf{x}}_{l}$ ($\mathbf{\hat{x}}_{l}$ ) is \emph{%
contained in} (\emph{or at most coincides with}) $\mathbf{\hat{x}}_{l}$ ($%
\overset{%
\mathchoice{}{}{\mbox{\raisebox{-.5ex}[0pt][0pt]{$\scriptstyle
\smallfrown$}}}{\mbox{\raisebox{-.35ex}[0pt][0pt]{$\scriptscriptstyle
\smallfrown$}}}}{\mathbf{x}}_{l}$). This entails that: a) an \emph{overall
estimate} of the system state $\mathbf{x}_{l}$ can be generated on the basis
of the posterior pdfs of $\overset{%
\mathchoice{}{}{\mbox{\raisebox{-.5ex}[0pt][0pt]{$\scriptstyle
\smallfrown$}}}{\mbox{\raisebox{-.35ex}[0pt][0pt]{$\scriptscriptstyle
\smallfrown$}}}}{\mathbf{x}}_{l}$\ and\ $\mathbf{\hat{x}}_{l}$ evaluated by F%
$_{1}$ and F$_{2}$, respectively; b) the portion $[x_{D-\hat{D},l},x_{D-\hat{%
D}+1,l},...,x_{\overset{%
\mathchoice{}{}{\mbox{\raisebox{-.5ex}[0pt][0pt]{$\scriptstyle
\smallfrown$}}}{\mbox{\raisebox{-.35ex}[0pt][0pt]{$\scriptscriptstyle
\smallfrown$}}}}{D}-1,l}]^{T}$ of $\mathbf{x}_{l}$, consisting of 
\begin{equation}
N_{d}\triangleq \overset{%
\mathchoice{}{}{\mbox{\raisebox{-.5ex}[0pt][0pt]{$\scriptstyle
\smallfrown$}}}{\mbox{\raisebox{-.35ex}[0pt][0pt]{$\scriptscriptstyle
\smallfrown$}}}}{D}+\hat{D}-D  \label{redundancy_deg}
\end{equation}%
elements, is estimated by both F$_{1}$ and F$_{2}$. Consequently, rule R1
requires the parameter $N_{d}$ (\ref{redundancy_deg}), that represents the 
\emph{degree of redundancy} of the overall filtering algorithm, to be
strictly positive.

The \emph{second rule} (i.e., \textbf{R2}) has been inspired by the fact
that, generally speaking, iterative decoders of concatenated channel codes
are made of multiple SISO modules, one for each constituent code. The
implementation of this rule in TF requires accurately defining the \emph{%
nature} of the statistical information to be passed from each constituent
filter to the other one. Actually, this problem has been already tackled in
the development of MPF, where the information passed from a particle filter
to a bank of Kalman filters takes the form of \emph{pseudo-measurements}
(PMs) evaluated on the basis of the \emph{mathematical constraints}
established by state update equations \cite{Schon_2005}. The use of PMs
allows us to exploit the \emph{memory} characterizing the time evolution of
dynamic models (and representing the second fundamental property of each
\emph{error correction method} employed in digital communications).
Moreover, PMs can be processed as they were \emph{real measurements} \cite%
{Schon_2005}; for this reason, their use can be incorporated in the FG\
shown in Fig. \ref{Fig_1} by including a new MU, i.e. by adding a new
equality node through which the message emerging from the first MU (i.e.,
from the MU based on real measurements) is merged with a\ message conveying PM
information. This idea is implemented in the graphical model\footnote{%
Note that \emph{oriented} edges are used in our graphical models wherever
message passing along such edges can be accomplished along a single
direction only.} shown in Fig. \ref{Fig_2}-(b) and providing a detailed
description of the \emph{overall processing} accomplished by a SISO module
based on F$_{1}$ (a similar model can be easily drawn for F$_{2}$ by
interchanging the couple $(\overset{%
\mathchoice{}{}{\mbox{\raisebox{-.5ex}[0pt][0pt]{$\scriptstyle
\smallfrown$}}}{\mbox{\raisebox{-.35ex}[0pt][0pt]{$\scriptscriptstyle
\smallfrown$}}}}{\mathbf{x}}_{l}$, $\overset{%
\mathchoice{}{}{\mbox{\raisebox{-.5ex}[0pt][0pt]{$\scriptstyle
\smallsmile$}}}{\mbox{\raisebox{-.35ex}[0pt][0pt]{$\scriptscriptstyle
\smallsmile$}}}}{\mathbf{x}}_{l})$ with $(\mathbf{\hat{x}}_{l}$, $\mathbf{%
\bar{x}}_{l})$ in that figure). In fact, this model represents the F$_{1}$
filtering algorithm (F$_{1}$ block), the \emph{conversion}\ of the
statistical information provided from F$_{2}$ into a form useful to F$_{1}$
(F$_{1}$-IN block) and the \emph{generation} of the statistical information
made available by F$_{1}$ to F$_{2}$ (F$_{1}$-OUT block). Its structure can
be explained as follows:

1. The algorithm employed by F$_{1}$ is based on the \emph{Markov model} $%
\tilde{f}(\overset{%
\mathchoice{}{}{\mbox{\raisebox{-.5ex}[0pt][0pt]{$\scriptstyle
\smallfrown$}}}{\mbox{\raisebox{-.35ex}[0pt][0pt]{$\scriptscriptstyle
\smallfrown$}}}}{\mathbf{x}}_{l+1}|\overset{%
\mathchoice{}{} {\mbox{\raisebox{-.5ex}[0pt][0pt]{$\scriptstyle
\smallfrown$}}} {\mbox{\raisebox{-.35ex}[0pt][0pt]{$\scriptscriptstyle
\smallfrown$}}}}{\mathbf{x}}_{l},\overset{%
\mathchoice{}{}{\mbox{\raisebox{-.5ex}[0pt][0pt]{$\scriptstyle
\smallsmile$}}}{\mbox{\raisebox{-.35ex}[0pt][0pt]{$\scriptscriptstyle
\smallsmile$}}}}{\mathbf{x}}_{l})$ and on the \emph{observation model} $%
\tilde{f}(\mathbf{y}_{l}|\overset{%
\mathchoice{}{}{\mbox{\raisebox{-.5ex}[0pt][0pt]{$\scriptstyle
\smallfrown$}}}{\mbox{\raisebox{-.35ex}[0pt][0pt]{$\scriptscriptstyle
\smallfrown$}}}}{\mathbf{x}}_{l},\overset{%
\mathchoice{}{} {\mbox{\raisebox{-.5ex}[0pt][0pt]{$\scriptstyle
\smallsmile$}}} {\mbox{\raisebox{-.35ex}[0pt][0pt]{$\scriptscriptstyle
\smallsmile$}}}}{\mathbf{x}}_{l})$, that represent the \emph{exact} models $%
f(\overset{%
\mathchoice{}{} {\mbox{\raisebox{-.5ex}[0pt][0pt]{$\scriptstyle
\smallfrown$}}} {\mbox{\raisebox{-.35ex}[0pt][0pt]{$\scriptscriptstyle
\smallfrown$}}}}{\mathbf{x}}_{l+1}|\overset{%
\mathchoice{}{}{\mbox{\raisebox{-.5ex}[0pt][0pt]{$\scriptstyle
\smallfrown$}}}{\mbox{\raisebox{-.35ex}[0pt][0pt]{$\scriptscriptstyle
\smallfrown$}}}}{\mathbf{x}}_{l},\overset{%
\mathchoice{}{}{\mbox{\raisebox{-.5ex}[0pt][0pt]{$\scriptstyle
\smallsmile$}}}{\mbox{\raisebox{-.35ex}[0pt][0pt]{$\scriptscriptstyle
\smallsmile$}}}}{\mathbf{x}}_{l})$ and $f(\mathbf{y}_{l}|\overset{%
\mathchoice{}{} {\mbox{\raisebox{-.5ex}[0pt][0pt]{$\scriptstyle
\smallfrown$}}} {\mbox{\raisebox{-.35ex}[0pt][0pt]{$\scriptscriptstyle
\smallfrown$}}}}{\mathbf{x}}_{l},\overset{%
\mathchoice{}{}{\mbox{\raisebox{-.5ex}[0pt][0pt]{$\scriptstyle
\smallsmile$}}}{\mbox{\raisebox{-.35ex}[0pt][0pt]{$\scriptscriptstyle
\smallsmile$}}}}{\mathbf{x}}_{l})$, respectively, or \emph{approximations}
of one or both of them (as required by the \emph{third rule}, i.e. by 
\textbf{R3}). The pdf of the state component $\overset{%
\mathchoice{}{}{\mbox{\raisebox{-.5ex}[0pt][0pt]{$\scriptstyle
\smallsmile$}}}{\mbox{\raisebox{-.35ex}[0pt][0pt]{$\scriptscriptstyle
\smallsmile$}}}}{\mathbf{x}}_{l}$ (unknown to F$_{1}$) is provided by F$_{2}$
through the message $\vec{m}_{fe2}(\overset{%
\mathchoice{}{}{\mbox{\raisebox{-.5ex}[0pt][0pt]{$\scriptstyle
\smallsmile$}}}{\mbox{\raisebox{-.35ex}[0pt][0pt]{$\scriptscriptstyle
\smallsmile$}}}}{\mathbf{x}}_{l})$. Morever, as already stated above, the
forward estimate of $\overset{%
\mathchoice{}{} {\mbox{\raisebox{-.5ex}[0pt][0pt]{$\scriptstyle
\smallfrown$}}} {\mbox{\raisebox{-.35ex}[0pt][0pt]{$\scriptscriptstyle
\smallfrown$}}}}{\mathbf{x}}_{l}$ is computed by F$_{1}$ in two distinct MU
steps, the first one involving the message $\vec{m}_{ms}(\overset{%
\mathchoice{}{} {\mbox{\raisebox{-.5ex}[0pt][0pt]{$\scriptstyle
\smallfrown$}}} {\mbox{\raisebox{-.35ex}[0pt][0pt]{$\scriptscriptstyle
\smallfrown$}}}}{\mathbf{x}}_{l})$ (based on the \emph{measurement} $\mathbf{%
y}_{l}$), the second one involving the message $\vec{m}_{pm}(\overset{%
\mathchoice{}{} {\mbox{\raisebox{-.5ex}[0pt][0pt]{$\scriptstyle
\smallfrown$}}} {\mbox{\raisebox{-.35ex}[0pt][0pt]{$\scriptscriptstyle
\smallfrown$}}}}{\mathbf{x}}_{l})$ (conveying the PM information computed by
F$_{2}$); these steps generate the messages $\vec{m}_{fe1}(\overset{%
\mathchoice{}{} {\mbox{\raisebox{-.5ex}[0pt][0pt]{$\scriptstyle
\smallfrown$}}} {\mbox{\raisebox{-.35ex}[0pt][0pt]{$\scriptscriptstyle
\smallfrown$}}}}{\mathbf{x}}_{l})$ and $\vec{m}_{fe2}(\overset{%
\mathchoice{}{} {\mbox{\raisebox{-.5ex}[0pt][0pt]{$\scriptstyle
\smallfrown$}}} {\mbox{\raisebox{-.35ex}[0pt][0pt]{$\scriptscriptstyle
\smallfrown$}}}}{\mathbf{x}}_{l})$, respectively.

2. The forward estimate $\vec{m}_{fe2}(\overset{%
\mathchoice{}{}{\mbox{\raisebox{-.5ex}[0pt][0pt]{$\scriptstyle
\smallfrown$}}}{\mbox{\raisebox{-.35ex}[0pt][0pt]{$\scriptscriptstyle
\smallfrown$}}}}{\mathbf{x}}_{l})$ computed by F$_{1}$ is passed to F$_{2}$
together with the PM message $\vec{m}_{pm}(\overset{%
\mathchoice{}{}{\mbox{\raisebox{-.5ex}[0pt][0pt]{$\scriptstyle
\smallsmile$}}}{\mbox{\raisebox{-.35ex}[0pt][0pt]{$\scriptscriptstyle
\smallsmile$}}}}{\mathbf{x}}_{l})$. The last message is evaluated on the
basis of the messages $\vec{m}_{fe1}(\overset{%
\mathchoice{}{}{\mbox{\raisebox{-.5ex}[0pt][0pt]{$\scriptstyle
\smallfrown$}}}{\mbox{\raisebox{-.35ex}[0pt][0pt]{$\scriptscriptstyle
\smallfrown$}}}}{\mathbf{x}}_{l})$ and $\vec{m}_{fe2}(\overset{%
\mathchoice{}{}{\mbox{\raisebox{-.5ex}[0pt][0pt]{$\scriptstyle
\smallfrown$}}}{\mbox{\raisebox{-.35ex}[0pt][0pt]{$\scriptscriptstyle
\smallfrown$}}}}{\mathbf{x}}_{l})$, i.e. on the basis of the forward
estimates available \emph{before} and \emph{after} the second MU of F$_{1}$.
Note also that the computation of $\vec{m}_{pm}(\overset{%
\mathchoice{}{}{\mbox{\raisebox{-.5ex}[0pt][0pt]{$\scriptstyle
\smallsmile$}}}{\mbox{\raisebox{-.35ex}[0pt][0pt]{$\scriptscriptstyle
\smallsmile$}}}}{\mathbf{x}}_{l})$ is carried out in the block called PM%
\emph{\ generation} (PMG) inside the F$_{1}$-OUT block.

3. The statistical information made available by F$_{2}$ to F$_{1}$ is
condensed in the messages $\vec{m}_{fe2}(\mathbf{\hat{x}}_{l})$ and $\vec{m}%
_{pm}(\mathbf{\bar{x}}_{l})$. The message $\vec{m}_{fe2}(\overset{%
\mathchoice{}{}{\mbox{\raisebox{-.5ex}[0pt][0pt]{$\scriptstyle
\smallsmile$}}}{\mbox{\raisebox{-.35ex}[0pt][0pt]{$\scriptscriptstyle
\smallsmile$}}}}{\mathbf{x}}_{l})$ acquired by F$_{1}$ can be computed by 
\emph{marginalizing} the message $\vec{m}_{fe2}(\mathbf{\hat{x}}_{l})$,
since, generally speaking, $\overset{%
\mathchoice{}{}{\mbox{\raisebox{-.5ex}[0pt][0pt]{$\scriptstyle
\smallsmile$}}}{\mbox{\raisebox{-.35ex}[0pt][0pt]{$\scriptscriptstyle
\smallsmile$}}}}{\mathbf{x}}_{l}$ is a portion of $\mathbf{\hat{x}}_{l}$
(marginalization is accomplished in block labelled with the letter M in Fig. %
\ref{Fig_2}-(b)); moreover, $\vec{m}_{fe2}(\mathbf{\hat{x}}_{l})$ is
processed jointly with $\vec{m}_{pm}(\mathbf{\bar{x}}_{l})$ to generate the
PM message $\vec{m}_{pm}(\overset{%
\mathchoice{}{}{\mbox{\raisebox{-.5ex}[0pt][0pt]{$\scriptstyle
\smallfrown$}}}{\mbox{\raisebox{-.35ex}[0pt][0pt]{$\scriptscriptstyle
\smallfrown$}}}}{\mathbf{x}}_{l})$ (this is accomplished in the block called
PM\emph{\ conversion}, PMC, inside the F$_{1}$-IN block).

\begin{figure}[tbp]
\centering
\includegraphics[width=0.8%
\textwidth]{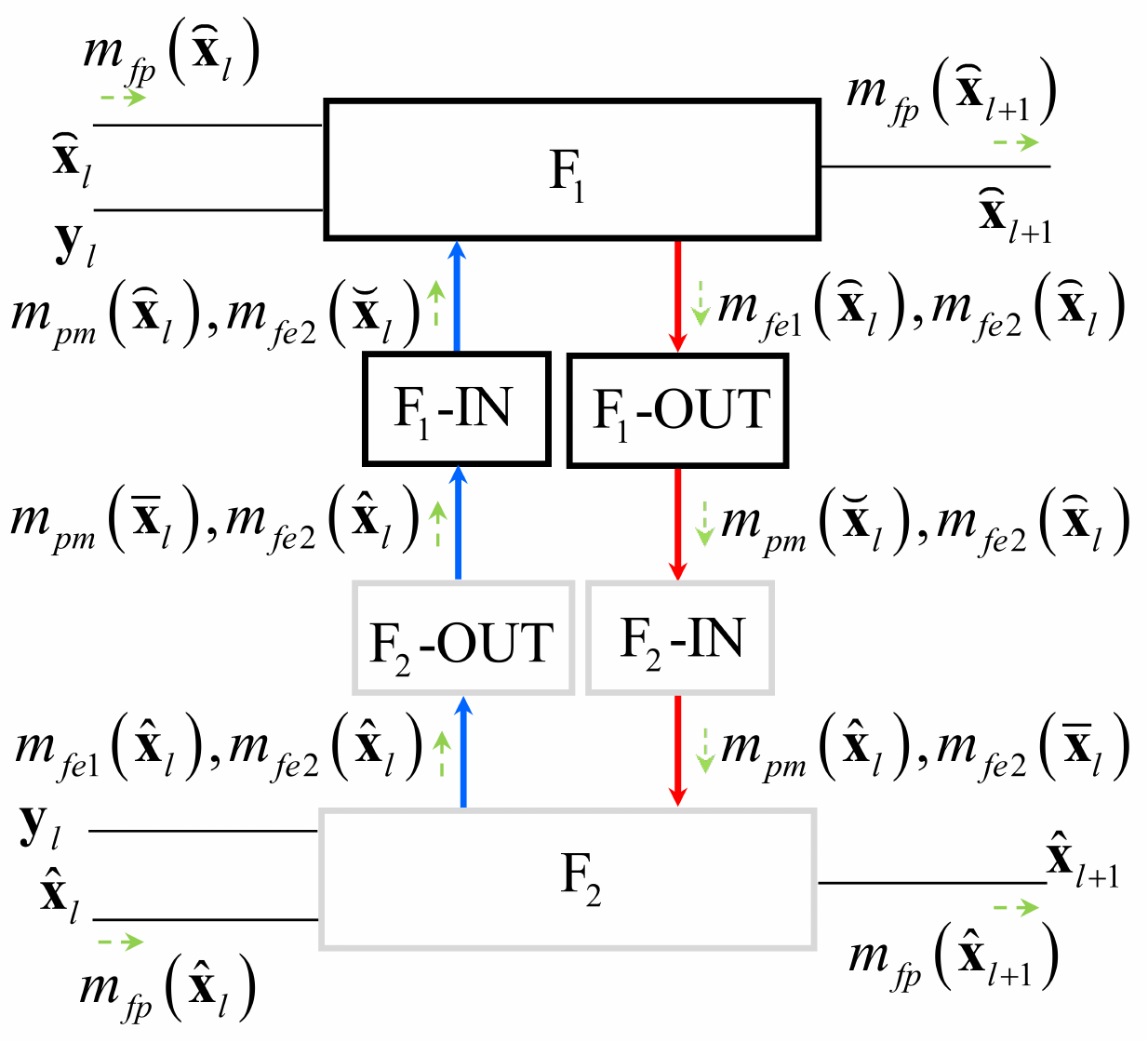}
\caption{Parallel concatenation of SISO modules based on filters F$_{1}$
and F$_{2}$; the flow of the messages exchanged between them is indicated by
green arrows.}
\label{Fig_3}
\end{figure}

Merging the graphical model shown in Fig. \ref{Fig_2}-(b) with its
counterpart referring to F$_{2}$ results in the PC architecture shown in
Fig. \ref{Fig_3}. This model, unlike the one illustrated in Fig. \ref{Fig_1}%
, is \emph{not cycle free}. For this reason, generally speaking, the
application of the SPA to it leads to \emph{iterative{} algorithms} with no
natural termination and whose accuracy can be substantially influenced by
the adopted \emph{message scheduling} \cite{Loeliger_2007}, \cite%
{Kschischang_2001}. This consideration and the possibility of choosing
different options for F$_{1}$ and F$_{2}$ lead easily to the conclusion that
the graphical models shown in Figs. \ref{Fig_2}-(b) and \ref{Fig_3} can be
employed to develop an entire \emph{family} of filtering algorithms, called 
\emph{turbo filters}.

\begin{figure}[tbp]
\centering
\includegraphics[width=0.8%
\textwidth]{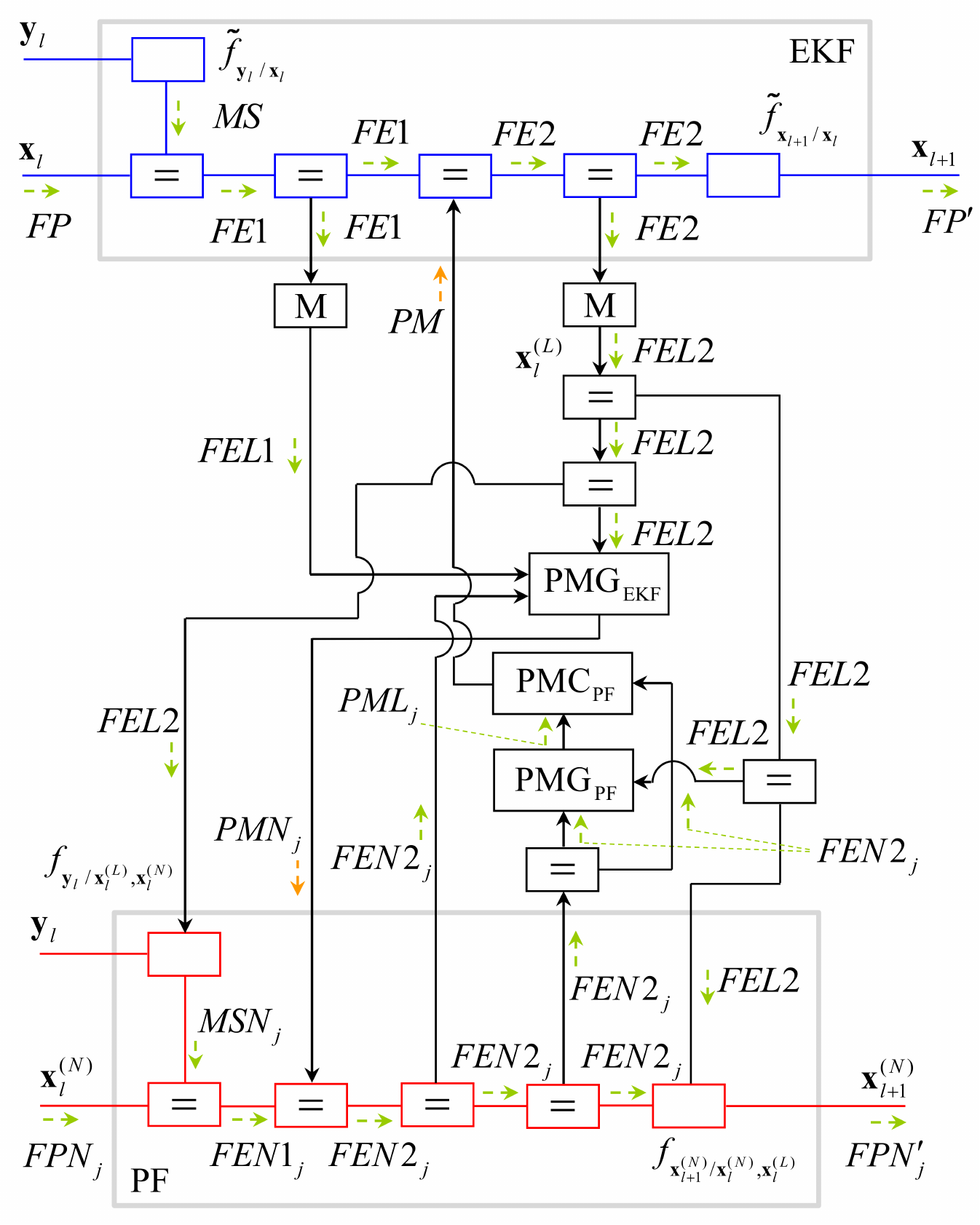}
\caption{Parallel concatenation of an extended Kalman filter with a particle
filter.}
\label{Fig_4}
\end{figure}

In the remaining part of this manuscript we focus on a specific instance of
the proposed PC architecture, since we make specific choices for both the
SSM and the two filters. In particular, we focus on the CLG\ SSM described in
Section \ref{sec:scenario} and assume that F$_{1}$ is an\emph{\ extended
Kalman filter} operating over the whole system state (so that $\overset{%
\mathchoice{}{} {\mbox{\raisebox{-.5ex}[0pt][0pt]{$\scriptstyle
\smallfrown$}}} {\mbox{\raisebox{-.35ex}[0pt][0pt]{$\scriptscriptstyle
\smallfrown$}}}}{\mathbf{x}}_{l}=\mathbf{x}_{l}$ and $\overset{%
\mathchoice{}{} {\mbox{\raisebox{-.5ex}[0pt][0pt]{$\scriptstyle
\smallsmile$}}} {\mbox{\raisebox{-.35ex}[0pt][0pt]{$\scriptscriptstyle
\smallsmile$}}}}{\mathbf{x}}_{l}$ is an empty vector), whereas \ F$_{2}$ is
a \emph{particle filter} (in particular, a \emph{sequential importance
resampling}, SIR, filter \cite{Arulampalam_2002}) operating on the nonlinear
state component only (so that $\mathbf{\hat{x}}_{l}=\mathbf{x}_{l}^{(N)}$
and $\mathbf{\bar{x}}_{l}=\mathbf{x}_{l}^{(L)}$); note that, in this case,
the degree of redundancy is $N_{d}=D_{N}$ (see (\ref{redundancy_deg})). Our
choices aim at developing a new concatenated filtering algorithm in which an
extended Kalman filter is aided by a particle filter in its most difficult
task, i.e. in the estimation of the nonlinear state component. Moreover, the
proposed TF scheme can be easily related to MPF, since the last technique can be
considered as a form of \emph{serial concatenation} of PF with Kalman
filtering. However, our TF instance employs, unlike MPF, a \emph{single}
(extended) Kalman filter in place of a bank of Kalman filters; morever, such
a filter estimates the whole system state, instead of its nonlinear
component only.
Based on the general models shown in Figs. \ref{Fig_2}-(b)
and \ref{Fig_3}, the specific graphical model illustrated in Fig. \ref{Fig_4}
can be drawn for the considered case. This model deserves the following
comments:

1. The upper (lower) rectangle delimited by a grey line allow to easily
identify the message passing accomplished by EKF (PF).

2. Filter F$_{1}$ is based on the \emph{approximate} models $\tilde{f}(%
\mathbf{x}_{l+1}|\mathbf{x}_{l})$ and $\tilde{f}(\mathbf{y}_{l}|\mathbf{x}%
_{l})$, that can be easily derived from the linearised eqs. (\ref%
{state_up_approx}) and (\ref{meas_mod_approx}), respectively. Moreover, the
(Gaussian) messages processed by it are $\vec{m}_{fp}(\mathbf{x}_{l})$, $%
\vec{m}_{ms}(\mathbf{x}_{l})$, $\vec{m}_{fe1}(\mathbf{x}_{l})$, $\vec{m}%
_{pm}(\mathbf{x}_{l})$, $\vec{m}_{fe2}(\mathbf{x}_{l})$ and $\vec{m}_{fp}(%
\mathbf{x}_{l+1})$, and are denoted $FP$, $MS$, $FE1$, $PM$, $FE2$ and $%
FP^{^{\prime }}$, respectively, to ease reading.

3. Filter F$_{2}$ is based on the \emph{exact} models $f(\mathbf{x}%
_{l+1}^{(N)}|\mathbf{x}_{l}^{(N)}$, $\mathbf{x}_{l}^{(L)})$ and $f(\mathbf{y}%
_{l}|\mathbf{x}_{l}^{(N)},\mathbf{x}_{l}^{(L)})$, that can be easily derived
from the eqs. (\ref{eq:XL_update}) (with $Z=N$) and (\ref{eq:y_t}),
respectively. Moreover, the messages processed by it and appearing in Fig. %
\ref{Fig_4} refer to the $j$-th particle \emph{predicted} in the previous
(i.e. $(l-1)$-th) recursion and denoted $\mathbf{x}_{fp,l,j}^{(N)}$, with $%
j=0,1,...,N_{p}-1$ (where $N_{p}$ represents the overall number of
particles); such messages are $\vec{m}_{fp,j}(\mathbf{x}_{l}^{(N)})$, $\vec{m%
}_{ms,j}(\mathbf{x}_{l}^{(N)})$, $\vec{m}_{fe1,j}(\mathbf{x}_{l}^{(N)})$, $%
\vec{m}_{pm,j}(\mathbf{x}_{l}^{(N)})$, $\vec{m}_{fe2,j}(\mathbf{x}%
_{l}^{(N)}) $ and $\vec{m}_{fp,j}(\mathbf{x}_{l+1}^{(N)})$, and are denoted $%
FPN_{j}$, $MSN_{j}$, $FEN1_{j}$, $PMN_{j}$, $FEN2_{j}$ and $%
FPN_{j}^{^{\prime }}$, respectively, to ease reading.

4. The message $\vec{m}_{fe1}(\mathbf{x}_{l})$ ($\vec{m}_{fe2}(\mathbf{x}%
_{l})$) generated by F$_{1}$ undergoes \emph{marginalization} in the block
labelled with the letter M; this results in the message $\vec{m}_{fe1}(\mathbf{%
x}_{l}^{(L)})$ ($\vec{m}_{fe2}(\mathbf{x}_{l}^{(L)})$), denoted $FEL1$ ($%
FEL2 $). Based on the general model shown in Fig. \ref{Fig_2}-b), we exploit
the messages $\vec{m}_{fe1}(\mathbf{x}_{l}^{(L)})$ and $\vec{m}_{fe2}(%
\mathbf{x}_{l}^{(L)})$ to compute the PM message $\vec{m}_{pm,j}(\mathbf{x}%
_{l}^{(N)})$ (denoted $PMN_{j}$) in the block called PMG$_{\text{EKF}}$.
Moreover, $\vec{m}_{fe2}(\mathbf{x}_{l}^{(L)})$ is employed for
marginalising the PF state
update and measurement models (i.e., $f(\mathbf{x}_{l+1}^{(N)}|\mathbf{x}%
_{l}^{(N)}$, $\mathbf{x}_{l}^{(L)})$ and  $f(\mathbf{y}_{l}|\mathbf{x}%
_{l}^{(N)},\mathbf{x}_{l}^{(L)})$, respectively); this allows us to compute
the messages $\vec{m}_{ms,j}(\mathbf{x}_{l}^{(N)})$ and $\vec{m}_{fp,j}(%
\mathbf{x}_{l+1}^{(N)})$, respectively.

5. The message $\vec{m}_{fe2,j}(\mathbf{x}_{l}^{(N)})$ produced by PF is
processed in the block called PMG$_{\text{PF}}$ in order to generate the PM
message $\vec{m}_{pm,j}(\mathbf{x}_{l}^{(L)})$ (the message $\vec{m}_{fe1,j}(%
\mathbf{x}_{l}^{(N)})$ is not required in this case; see the next Section).
Moreover, the two sets $\{\vec{m}_{pm,j}(\mathbf{x}_{l}^{(L)})\}$ and $\{%
\vec{m}_{fe2,j}(\mathbf{x}_{l}^{(N)})\}$ (each consisting of $N_{p}$
messages) are merged in the block called PMC$_{\text{PF}}$, where the
information they convey are \emph{converted} into the (single) PM message $%
\vec{m}_{pm}(\mathbf{x}_{l})$ feeding F$_{1}$.

6. At the end of the $l$-th recursion, a single statistical model is
available for $\mathbf{x}_{l}^{(L)}$. On the contrary, two models are
available for $\mathbf{x}_{l}^{(N)}$, one particle-based, the other one
Gaussian, since this state component is shared by F$_{1}$ and F$_{2}$; note
that the former model, unlike the second one, is able to represent a \emph{%
multimodal} pdf.

Let us now focus on the evaluation of the PMs for the considered TF scheme.
On the one hand, the PM messages $\{\vec{m}_{pm,j}(\mathbf{x}_{l}^{(N)})\}$
evaluated for F$_{2}$ are exploited to improve the estimation accuracy for
the \emph{nonlinear state component only}. Their computation involves the
pdf of the random vector 
\begin{equation}
\mathbf{z}_{l}^{(N)}\triangleq \mathbf{x}_{l+1}^{(L)}-\mathbf{A}%
_{l}^{(L)}\left( \mathbf{x}_{l}^{(N)}\right) \mathbf{x}_{l}^{(L)}\text{,}
\label{z_N_l}
\end{equation}%
defined on the basis of the state update equation (\ref{eq:XL_update}) (with 
$Z=L$). This pdf need to be evaluated for each of the $N_{p}$ particles
representing $\mathbf{x}_{l}^{(N)}$; in the following, its expression
associated with the $j$-th particle (i.e., conditioned on $\mathbf{x}%
_{l}^{(N)}=\mathbf{x}_{fp,l,j}^{(N)}$) and evaluated on the basis of the
joint pdf of $\mathbf{x}_{l}^{(L)}$ and $\mathbf{x}_{l+1}^{(L)}$ provided by
F$_{1}$ is conveyed by the message $\vec{m}_{j}(\mathbf{z}_{l}^{(N)})$. Note
also that, based on (\ref{eq:XL_update}) (with $Z=L$), the vector $\mathbf{z}%
_{l}^{(N)}$ (\ref{z_N_l}) is expected to equal the sum 
\begin{equation}
\mathbf{f}_{l}^{(L)}\left( \mathbf{x}_{l}^{(N)}\right) +\mathbf{w}_{l}^{(L)},
\label{z_N_l_bis}
\end{equation}%
that depends on $\mathbf{x}_{l}^{(N)}$ only; the pdf of $\mathbf{z}%
_{l}^{(N)} $ evaluated on the basis of (\ref{z_N_l_bis}) is denoted $f(%
\mathbf{z}_{l}^{(N)}|\mathbf{x}_{l}^{(N)})$ in the following.

On the other hand, the PM message $\vec{m}_{pm}(\mathbf{x}_{l})$ evaluated
for F$_{1}$ is expected to improve the estimation accuracy for the \emph{%
whole state}. For this reason, in our TF techniques, its computation involves
the two message sets $\{\vec{m}_{pm,j}(\mathbf{x}_{l}^{(L)})\}$ and $\{\vec{m%
}_{fe2,j}(\mathbf{x}_{l}^{(N)})\}$, generated by F$_{2}$ and referring to
the two distinct components of $\mathbf{x}_{l}$. The messages $\{\vec{m}%
_{fe2,j}(\mathbf{x}_{l}^{(N)})\}$ convey a particle-based representation of $%
\mathbf{x}_{l}^{(N)}$. The message $\vec{m}_{pm,j}(\mathbf{x}_{l}^{(L)})$,
instead, represents the pdf of the random vector \cite{Schon_2005}%
\begin{equation}
\mathbf{z}_{l}^{(L)}\triangleq \mathbf{x}_{l+1}^{(N)}-\mathbf{f}%
_{l}^{(N)}\left( \mathbf{x}_{l}^{(N)}\right)  \label{eq:z_L_l}
\end{equation}%
conditioned on $\mathbf{x}_{l}^{(N)}=\mathbf{x}_{fp,l,j}^{(N)}$ for any $j$.
This pdf is evaluated on the basis of the joint representation of the couple 
$(\mathbf{x}_{l}^{(N)}$, $\mathbf{x}_{l+1}^{(N)})$ produced by F$_{2}$ and
is conveyed by the message $\vec{m}_{j}(\mathbf{z}_{l}^{(L)})$; note also
that, based on (\ref{eq:XL_update}) (with $Z=N$), the quantity $\mathbf{z}%
_{l}^{(L)}$ (\ref{eq:z_L_l}) is expected to equal the sum

\begin{equation}
\mathbf{A}_{l}^{(N)}\left( \mathbf{x}_{l}^{(N)}\right) \mathbf{x}_{l}^{(L)}+%
\mathbf{w}_{l}^{(N)},  \label{eq:z_L_l_bis}
\end{equation}%
that depends on $\mathbf{x}_{l}^{(L)}$ and $\mathbf{x}_{l}^{(N)}$ only; the
pdf of $\mathbf{z}_{l}^{(N)}$ evaluated on the basis of (\ref{eq:z_L_l_bis})
is denoted $f(\mathbf{z}_{l}^{(L)}|\mathbf{x}_{l}^{(L)},\mathbf{x}%
_{l}^{(N)}) $ in the following.

Two specific message scheduling for the graphical model shown in Fig. \ref%
{Fig_4} are proposed in the following Section, where the computation of all
the involved messages is also analysed in detail.

\section{Message Passing in Turbo Filtering\label{sec:Message-Passing}}

\begin{figure}[tbp]
\centering
\includegraphics[width=0.8%
\textwidth]{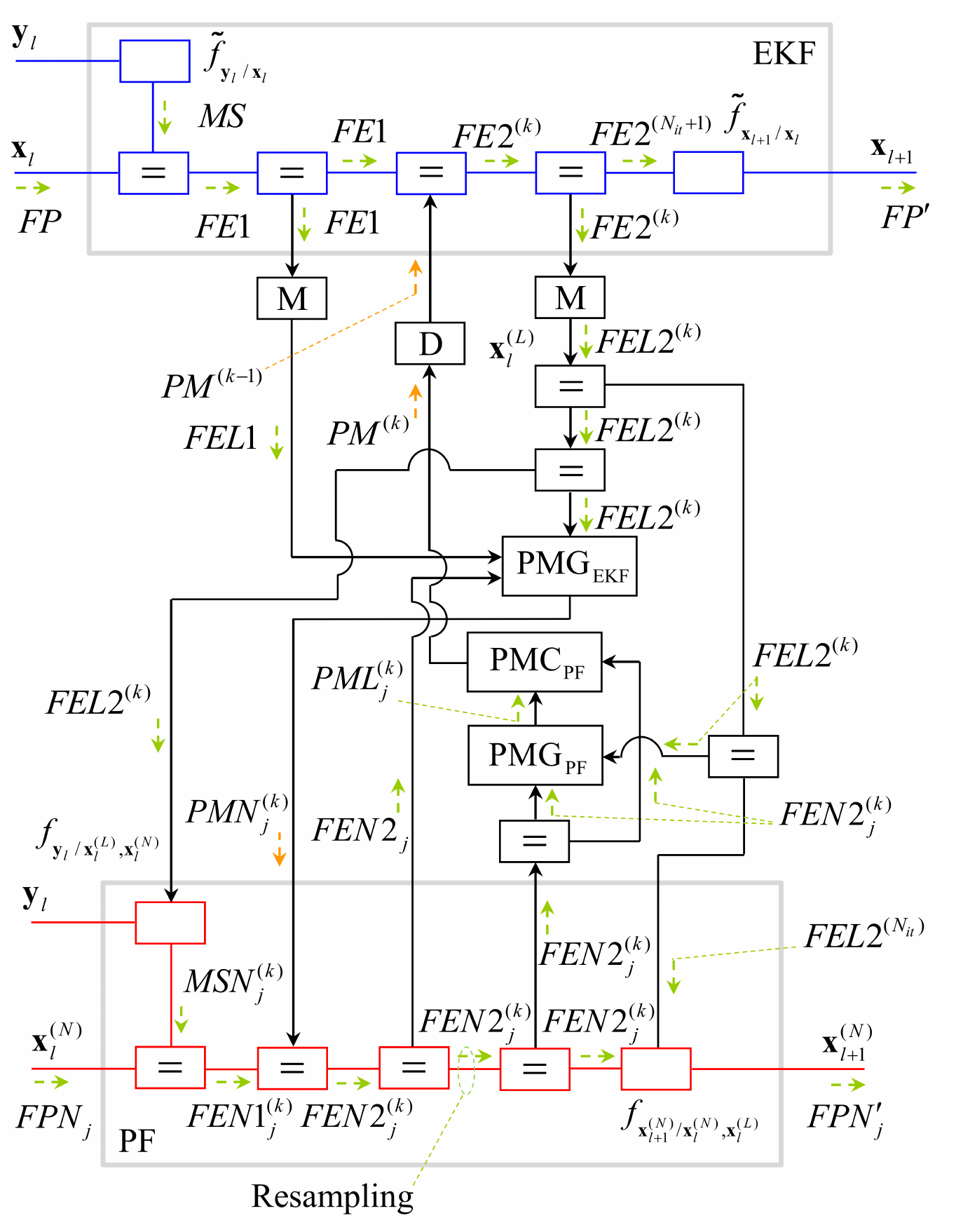}
\caption{Message scheduling adopted in TF\#1.}
\label{Fig_5}
\end{figure}

In this Section two different options are considered for the scheduling of
the messages appearing in Fig. \ref{Fig_4}. The first option consists in
running EKF before PF within each iteration, whereas the second one in doing
the opposite; the resulting algorithms are dubbed TF\#1 and TF\#2,
respectively. The message scheduling adopted in TF\#1 is represented in
Fig. \ref{Fig_5}, that refers to the $k$-th iteration accomplished within
the $l$-th recursion (with $k=1,2,...,N_{it}$, where $N_{it}$ is the overall
number of iterations); this explains why the superscripts $(k)$ \ and $(k-1)$
have been added to all the iteration-dependent messages appearing in Fig. %
\ref{Fig_4}.\ 

As far as the evaluation of the messages passed in TF\#1 and TF\#2 is
concerned, this is mainly based on three \emph{computational rules} (CR)
resulting from the application of the SPA to equality nodes and function
nodes. More specifically, the first computational rule, denoted CR1, applies
to an \emph{equality constraint node}; if the messages $\vec{m}_{1}\left( 
\mathbf{x}\right) $ and $\vec{m}_{2}\left( \mathbf{x}\right) $ denote the
messages entering it, the message $\vec{m}_{3}\left( \mathbf{x}\right) =\vec{%
m}_{1}\left( \mathbf{x}\right) \vec{m}_{2}\left( \mathbf{x}\right) $ emerges
from it. In particular, if $\vec{m}_{i}\left( \mathbf{x}\right) =\mathcal{N(}%
\mathbf{x};\mathbf{\eta }_{i},\mathbf{C}_{i})$ (with $i=1$ and $2$), then $%
\vec{m}_{3}\left( \mathbf{x}\right) =\mathcal{N(}\mathbf{x};\mathbf{\eta }%
_{3},\mathbf{C}_{3})$; moreover, the precision matrix $\mathbf{W}_{3}$ and
the transformed mean vector $\mathbf{w}_{3}$ associated with $\mathbf{C}_{3}$
and\ $\mathbf{\eta }_{3}$, respectively, are given by (see \cite[Table 2, p.
1303, eqs. (II.1) and (II.2)]{Loeliger_2007})%
\begin{equation}
\mathbf{W}_{3}\triangleq \mathbf{C}_{3}^{-1}=\mathbf{W}_{1}+\mathbf{W}_{2}
\label{Prec_mat}
\end{equation}%
and 
\begin{equation}
\mathbf{w}_{3}\triangleq \mathbf{C}_{3}^{-1}\mathbf{\eta }_{3}=\mathbf{w}%
_{1}+\mathbf{w}_{2}  \label{Transf_mean_vector}
\end{equation}%
respectively, where $\mathbf{W}_{i}\triangleq \mathbf{C}_{i}^{-1}$ and $%
\mathbf{w}_{i}\triangleq \mathbf{C}_{i}^{-1}\mathbf{\eta }_{i}$ for $i=1$, $%
2 $. The second computational rule, denoted CR2, applies to a node
representing the function $f\left( \mathbf{x}_{1},\mathbf{x}_{2}\right) $;
if the message $\vec{m}_{1}\left( \mathbf{x}_{1}\right) $ denotes the
message entering it, the message $\vec{m}_{2}\left( \mathbf{x}_{2}\right) $
emerging from it is given by 
\begin{equation}
\vec{m}_{2}\left( \mathbf{x}_{2}\right) =\int \vec{m}_{1}\left( \mathbf{x}%
_{1}\right) f\left( \mathbf{x}_{1},\mathbf{x}_{2}\right) d\mathbf{x}_{1}.
\label{funct_rule}
\end{equation}%
In particular, if $\vec{m}_{1}\left( \mathbf{x}_{1}\right) =\mathcal{N(}%
\mathbf{x}_{1};\mathbf{\eta }_{1},\mathbf{C}_{1})$ and $f\left( \mathbf{x}%
_{1},\mathbf{x}_{2}\right) =\mathcal{N}\left( \mathbf{x}_{2};\mathbf{Ax}_{1}+%
\mathbf{b},\mathbf{C}\right) $, then%
\begin{equation}
\vec{m}_{2}\left( \mathbf{x}_{2}\right) =\mathcal{N(}\mathbf{x}_{2};\mathbf{%
\eta }_{2},\mathbf{C}_{2}),  \label{m_2}
\end{equation}%
with $\mathbf{\eta }_{2}=\mathbf{A\eta }_{1}+\mathbf{b}$ and $\mathbf{C}_{2}=%
\mathbf{C}+\mathbf{AC}_{1}\left( \mathbf{A}\right) ^{T}$ (see \cite[Table 2,
p. 1303, eqs. (II.7) and (II.9); Table 3, p. 1304, eqs. (III.1) and (III.3) ]%
{Loeliger_2007}). Finally, the third computational rule, denoted CR3,
applies to a node representing the function $f\left( \mathbf{x}\right) =%
\mathcal{N}(\mathbf{x};\mathbf{\eta }_{2},\mathbf{C}_{2})$ and fed by the
message $\vec{m}_{1}\left( \mathbf{x}\right) =\mathcal{N(}\mathbf{x};\mathbf{%
\eta }_{1},\mathbf{C}_{1})$; the output message is the \emph{constant}
message 
\begin{equation}
\vec{m}_{2}=D\,\exp \left[ \frac{1}{2}\left( \mathbf{\eta }^{T}\mathbf{W\eta 
}-\mathbf{\eta }_{1}^{T}\mathbf{W}_{1}\mathbf{\eta }_{1}-\mathbf{\eta }%
_{2}^{T}\mathbf{W}_{2}\mathbf{\eta }_{2}\right) \right]  \label{eq:CR3}
\end{equation}%
where $\mathbf{\mathbf{W}}_{1}\triangleq \mathbf{C}_{1}^{-1}$, $\mathbf{%
\mathbf{W}}_{2}\triangleq \mathbf{C}_{2}^{-1}$, $\mathbf{W}=\mathbf{\mathbf{W%
}}_{1}+\mathbf{\mathbf{W}}_{2}$, $\mathbf{W}\mathbf{\eta }=\mathbf{W}_{1}%
\mathbf{\eta }_{1}+\mathbf{W}_{2}\mathbf{\eta }_{2}$, $D=\left( \det \left[ 
\mathbf{C}_{1}+\mathbf{C}_{2}\right] \right) ^{-N/2}$ and $N$ is the size of 
$\mathbf{x}$.

In the following we show how, applying the above mentioned CRs, simple
formulas can be derived for all messages passed in the graphical model shown
in Fig. \ref{Fig_5}. However, before doing this, we need to define the \emph{%
input} messages for the considered recursion; these are%
\begin{equation}
\vec{m}_{fp}\left( \mathbf{x}_{l}\right) =\mathcal{N}\left( \mathbf{x}_{l};%
\mathbf{\eta }_{fp,l},\mathbf{C}_{fp,l}\right)
\label{eq:eq:message_fp_l_EKF}
\end{equation}%
for the EKF (upper part of the graphical model)\ and the set of $N_{p}$
messages $\{\vec{m}_{fp,j}(\mathbf{x}_{l}^{(N)})\}$ for the PF (lower part
of the graphical model), where 
\begin{equation}
\vec{m}_{fp,j}\left( \mathbf{x}_{l}^{(N)}\right) =\mathbf{\delta }\left( 
\mathbf{x}_{l}^{(N)}-\mathbf{x}_{fp,l,j}^{(N)}\right) ,
\label{eq:message_fp_l_j_PF}
\end{equation}%
with $j=0,1,...,N_{p}-1$; in the following we also assume that the $N_{p}$
available particles are collected in the set $S_{l}\triangleq \{\mathbf{x}%
_{fp,l,j}^{(N)}\}$. On the other hand, the \emph{output} messages are $\vec{m%
}_{fp}\left( \mathbf{x}_{l+1}\right) $ (for EKF) and $\{\vec{m}_{fp,j}(%
\mathbf{x}_{l+1}^{(N)})\}$ (for PF); since, as shown below, the devised TF
algorithms preserve the mathematical structure of the filtered densities
from recursion to recursion, $\vec{m}_{fp}\left( \mathbf{x}_{l+1}\right) $
and $\vec{m}_{fp,j}(\mathbf{x}_{l+1}^{(N)})$ have the same functional form
as $\vec{m}_{fp}(\mathbf{x}_{l})$ (\ref{eq:eq:message_fp_l_EKF}) and $\vec{m}%
_{fp,j}(\mathbf{x}_{l}^{(N)})$ (\ref{eq:message_fp_l_j_PF}) (for any $j$),
respectively.

It is also worth mentioning that not all the messages appearing in Fig. \ref%
{Fig_5} depend on the iteration index $k$. More specifically, the following
messages are computed only once:

1. The messages $\vec{m}_{fe1}\left( \mathbf{x}_{l}\right) $ and $\vec{m}%
_{fe1}(\mathbf{x}_{l}^{(L)})$ evaluated by EKF in its \emph{first} MU. In
particular, $\vec{m}_{fe1}\left( \mathbf{x}_{l}\right) $ is computed as (see
Fig. \ref{Fig_5})%
\begin{equation}
\vec{m}_{fe1}(\mathbf{x}_{l})=\vec{m}_{fp}(\mathbf{x}_{l})\,\vec{m}%
_{ms}\left( \mathbf{x}_{l}\right) ,  \label{m_fe1_EKF_bis}
\end{equation}%
where $\vec{m}_{ms}\left( \mathbf{x}_{l}\right) $ is the message conveying
the information provided by $\mathbf{y}_{l}$, whose statistical
representation is expressed by the pdf $\tilde{f}(\mathbf{y}_{l}|\mathbf{x}%
_{l})$ (resulting from the linearised equation (\ref{meas_mod_approx}));
therefore, it can be expressed as 
\begin{equation}
\vec{m}_{ms}\left( \mathbf{x}_{l}\right) =\mathcal{N}\left( \mathbf{y}_{l};%
\mathbf{H}_{l}^{T}\mathbf{x}_{l}+\mathbf{v}_{l},\mathbf{C}_{e}\right) \text{,%
}  \label{m_ms_EKF}
\end{equation}%
or, equivalently, as (see \cite[Table 3, p. 1304, eqs. (III.5) and (III.6) ]%
{Loeliger_2007}) 
\begin{equation}
\vec{m}_{ms}\left( \mathbf{x}_{l}\right) =\mathcal{N}\left( \mathbf{x}_{l};%
\mathbf{\eta }_{ms,l},\mathbf{C}_{ms,l}\right) ;  \label{m_ms_EKF_bis}
\end{equation}%
here, the covariance matrix $\mathbf{C}_{ms,l}$ \ and the mean vector $%
\mathbf{\eta }_{ms,l}$ can be evaluated from the associated precision matrix%
\begin{equation}
\mathbf{W}_{ms,l}\triangleq \left( \mathbf{C}_{ms,l}\right) ^{-1}=\mathbf{H}%
_{l}\mathbf{W}_{e}\mathbf{H}_{l}^{T},  \label{w_ms_EKF}
\end{equation}%
and the transformed mean vector%
\begin{equation}
\mathbf{w}_{ms,l}\triangleq \mathbf{W}_{ms,l}\mathbf{\eta }_{ms,l}=\mathbf{H}%
_{l}\mathbf{W}_{e}\left( \mathbf{y}_{l}-\mathbf{v}_{l}\right) ,
\label{W_ms_EKF}
\end{equation}%
respectively, and $\mathbf{W}_{e}\triangleq \mathbf{C}_{e}^{-1}$. Therefore, 
$\vec{m}_{fe1}\left( \mathbf{x}_{l}\right) $ (\ref{m_fe1_EKF_bis}) can be
put in the form%
\begin{equation}
\vec{m}_{fe1}\left( \mathbf{x}_{l}\right) =\mathcal{N}\left( \mathbf{x}_{l};%
\mathbf{\eta }_{fe1,l},\mathbf{C}_{fe1,l}\right) ,  \label{m_fe1_EKF}
\end{equation}%
where the covariance matrix $\mathbf{C}_{fe1,l}$ \ and the mean vector $%
\mathbf{\eta }_{fe1,l}$ can be evaluated from the associated precision
matrix (see CR1, eq. (\ref{Prec_mat})) 
\begin{equation}
\mathbf{W}_{fe1,l}\triangleq \left( \mathbf{C}_{fe1,l}\right) ^{-1}=\mathbf{W%
}_{fp,l}+\mathbf{W}_{ms,l}  \label{W_fe1_EKF}
\end{equation}%
and the transformed mean vector (see CR1, eq. (\ref{Transf_mean_vector})) 
\begin{equation}
\mathbf{w}_{fe1,l}\triangleq \mathbf{W}_{fe1,l}\mathbf{\eta }_{fe1,l}=%
\mathbf{w}_{fp,l}+\mathbf{w}_{ms,l},  \label{w_fe1_EKF}
\end{equation}%
respectively; here, $\mathbf{W}_{fp,l}\triangleq (\mathbf{C}_{fp,l})^{-1}$
and $\mathbf{w}_{fp,l}\triangleq \mathbf{W}_{fp,l}\mathbf{\eta }_{fp,l}$.
The message $\vec{m}_{fe1}(\mathbf{x}_{l}^{(L)})$, instead, is easily
obtained from $\vec{m}_{fe1}(\mathbf{x}_{l})$ (\ref{m_fe1_EKF}) by
marginalizing the last message with respect to $\mathbf{x}_{l}^{(N)}$; this
produces 
\begin{equation}
\vec{m}_{fe1}\left( \mathbf{x}_{l}^{(L)}\right) =\int \vec{m}_{fe1}\left( 
\mathbf{x}_{l}\right) d\mathbf{x}_{l}^{(N)}=\mathcal{N(}\mathbf{x}_{l}^{(L)};%
\mathbf{\tilde{\eta}}_{fe1,l},\mathbf{\tilde{C}}_{fe1,l}),
\label{m_fe_L_EKF_1}
\end{equation}%
where $\mathbf{\tilde{\eta}}_{fe1,l}$ and $\mathbf{\tilde{C}}_{fe1,l}$ are
extracted from the mean $\mathbf{\eta }_{fe1,l}$ and the covariance matrix $%
\mathbf{C}_{fe1,l}$ of $\vec{m}_{fe1}(\mathbf{x}_{l})$, respectively, since $%
\mathbf{x}_{l}^{(L)}$ consists of the first $D_{L}$ elements of $\mathbf{x}%
_{l}$.

2. The output messages $\vec{m}_{fp}\left( \mathbf{x}_{l+1}\right) $ and $%
\vec{m}_{fp,j}(\mathbf{x}_{l+1}^{(N)})$ (for any $j$), since they are
evaluated on the basis of the forward estimates $\vec{m}_{fe2}^{(N_{it})}%
\left( \mathbf{x}_{l}\right) $ and $\{\vec{m}_{fe2,j}^{(N_{it}+1)}(\mathbf{x}%
_{l}^{(N)})\}$ computed by EKF and PF, respectively, in the last iteration.

In the following, a detailed description of the messages passed in TF\#1 is
provided. The formulas derived for this algorithm can be easily re-used in
the computation the messages passed in TF\#2; for this reason, after
developing TF\#1, we limit to providing a brief description of the
scheduling adopted in TF\#2.

The scheduling illustrated in Fig. \ref{Fig_5} for TF\#1 consists in
computing the involved (iteration-dependent) messages according to the
following order: 1) $\vec{m}_{fe2}^{(k)}\left( \mathbf{x}_{l}\right) $, $%
\vec{m}_{fe2}^{(k)}(\mathbf{x}_{l}^{(L)})$; 2) $\{\vec{m}_{ms,j}^{(k)}(%
\mathbf{x}_{l}^{(N)})\}$, $\{\vec{m}_{fe1,j}^{(k)}(\mathbf{x}_{l}^{(N)})\}$;
3) $\{\vec{m}_{pm,j}^{(k)}(\mathbf{x}_{l}^{(N)})\}$, $\{\vec{m}%
_{fe2,j}^{(k)}(\mathbf{x}_{l}^{(N)})\}$; 4) $\{\vec{m}_{pm,j}^{(k)}(\mathbf{x%
}_{l}^{(L)})\}$, $\vec{m}_{pm}^{(k)}(\mathbf{x}_{l})$. Therefore, the
evaluation of these messages can be organized according to the four steps
described below and to be carried out for $k=1$, $2$, $...$, $N_{it}$. Note
that in our description of TF\#1 scheduling, particle-dependent messages
always refer to the $j$-th particle (with that $j=0,1,...,N_{p}-1$) and
that, generally speaking, the structure of the particle set changes from
iteration to iteration, even if it preserves its cardinality; moreover, the
particle set available at the beginning of the $k$-th iteration is $%
S_{l}^{(k-1)}=\{\mathbf{x}_{fp,l,j}^{(N)}[k-1],j=0,1,...,N_{p}-1\}$, with $%
S_{l}^{(0)}=S_{l}$ and $\mathbf{x}_{fp,l,j}^{(N)}[0]=\mathbf{x}%
_{fp,l,j}^{(N)}$.

1. \emph{Second }MU\emph{\ in }EKF - This step aims at updating our
statistical knowledge about $\mathbf{x}_{l}$ on the basis of the PM
information conveyed by the message $\vec{m}_{pm}^{(k-1)}(\mathbf{x}_{l})$
(computed in the previous iteration on the basis of the statistical
information generated by PF; see step 4.). This is carried out by computing
the new message (see Fig. \ref{Fig_5}) 
\begin{equation}
\vec{m}_{fe2}^{(k)}\left( \mathbf{x}_{l}\right) =\vec{m}_{pm}^{(k-1)}\left( 
\mathbf{x}_{l}\right) \,\vec{m}_{fe1}\left( \mathbf{x}_{l}\right) ,
\label{m_fe2_EKF}
\end{equation}%
where $\vec{m}_{fe1}\left( \mathbf{x}_{l}\right) $ is expressed by (\ref%
{m_fe1_EKF}), and $\vec{m}_{pm}^{(k-1)}(\mathbf{x}_{l})$ is equal to unity
for $k=1$ (because of the adopted scheduling) and is given by (\ref%
{message_pm_x_l_k}) for $k>1$. Consequently, 
\begin{equation}
\vec{m}_{fe2}^{(k)}\left( \mathbf{x}_{l}\right) =\mathcal{N}\left( \mathbf{x}%
_{l};\mathbf{\eta }_{fe2,l}^{(k)},\mathbf{C}_{fe2,l}^{(k)}\right) ,
\label{m_fe2_EKF_k}
\end{equation}%
where $\mathbf{\eta }_{fe2,l}^{(k)}=\mathbf{\eta }_{fe1,l}$ and $\mathbf{C}%
_{fe2,l}^{(k)}=\mathbf{C}_{fe1,l}$ for $k=1$, whereas, for $k>1$, the
covariance matrix $\mathbf{C}_{fe2}^{(k)}$ \ and the mean vector $\mathbf{%
\eta }_{fe2}^{(k)}$ are evaluated as (see CR1, eq. (\ref{Prec_mat})) 
\begin{equation}
\mathbf{C}_{fe2,l}^{(k)}=\mathbf{W}_{l}^{(k-1)}\mathbf{C}_{pm,l}^{(k-1)}
\label{W_fe2_EKF_k}
\end{equation}%
and (see CR1, eq. (\ref{Transf_mean_vector})) 
\begin{equation}
\mathbf{\eta }_{fe2,l}^{(k)}=\mathbf{W}_{l}^{(k-1)}\left[ \mathbf{C}%
_{pm,l}^{(k-1)}\mathbf{w}_{fe1,l}+\mathbf{\eta }_{pm,l}^{(k-1)}\right] ,
\label{w_fe2_EKF_k}
\end{equation}%
respectively; here, $\mathbf{W}_{l}^{(k-1)}\triangleq \lbrack \mathbf{C}%
_{pm,l}^{(k-1)}\mathbf{W}_{fe1,l}+\mathbf{I}_{D}]^{-1}$. Marginalizing the
message $\vec{m}_{fe2}^{(k)}\left( \mathbf{x}_{l}\right) $ (\ref{m_fe2_EKF_k}%
) with respect to $\mathbf{x}_{l}^{(N)}$ results in the message 
\begin{equation}
\vec{m}_{fe2}^{(k)}\left( \mathbf{x}_{l}^{(L)}\right) \triangleq \int \vec{m}%
_{fe2}^{(k)}\left( \mathbf{x}_{l}\right) d\mathbf{x}_{l}^{(N)}=\mathcal{N(}%
\mathbf{x}_{l}^{(L)};\mathbf{\tilde{\eta}}_{fe2,l}^{(k)},\mathbf{\tilde{C}}%
_{fe2,l}^{(k)}),  \label{m_fe_L_EKF_2}
\end{equation}%
where $\mathbf{\tilde{\eta}}_{fe2,l}^{(k)}$ and $\mathbf{\tilde{C}}%
_{fe2,l}^{(k)}$ are easily extracted from the mean $\mathbf{\eta }%
_{fe2,l}^{(k)}$ and the covariance matrix $\mathbf{C}_{fe2,l}^{(k)}$ of $%
\vec{m}_{fe2}^{(k)}\left( \mathbf{x}_{l}\right) $ (\ref{m_fe2_EKF_k}),
respectively, since $\mathbf{x}_{l}^{(L)}$ consists of the first $D_{L}$
elements of $\mathbf{x}_{l}$.

2. \emph{First }MU\emph{\ in} PF - This step aims at updating the weight of
the $j$-th particle $\mathbf{x}_{fp,l,j}^{(N)}[k-1]$, conveyed by the
message (see (\ref{eq:message_fp_l_j_PF}))%
\begin{equation}
\vec{m}_{fp,j}^{(k)}(\mathbf{x}_{l}^{(N)})=\mathbf{\delta }(\mathbf{x}%
_{l}^{(N)}-\mathbf{x}_{fp,l,j}^{(N)}[k-1]),  \label{eq:message_fp_l_j_PF_bis}
\end{equation}%
on the basis of the new measurements $\mathbf{y}_{l}$. It involves the
computation of the messages $\vec{m}_{ms,j}^{(k)}(\mathbf{x}_{l}^{(N)})$ and
(see Fig. \ref{Fig_5})%
\begin{equation}
\vec{m}_{fe1,j}^{(k)}\left( \mathbf{x}_{l}^{(N)}\right) =\vec{m}%
_{ms,j}^{(k)}\left( \mathbf{x}_{l}^{(N)}\right) \,\vec{m}_{fp,j}^{(k)}\left( 
\mathbf{x}_{l}^{(N)}\right) .  \label{m_fe1_k_PF}
\end{equation}%
The evaluation of the message $\vec{m}_{ms,j}^{(k)}(\mathbf{x}_{l}^{(N)})$
requires \emph{marginalizing} the measurement model $f(\mathbf{y}_{l}|%
\mathbf{x}_{l}^{(N)},\,\mathbf{x}_{l}^{(L)})$ with respect to $\,\mathbf{x}%
_{l}^{(N)}$ (see Fig. \ref{Fig_5}), whose pdf is provided by the message $%
\vec{m}_{fe2}^{(k)}(\mathbf{x}_{l}^{(L)})$ (\ref{m_fe_L_EKF_2}). Therefore,
the message $\vec{m}_{ms,j}^{(k)}(\mathbf{x}_{l}^{(N)})$ emerging from the
function node representing $f(\mathbf{y}_{l}|\mathbf{x}_{l}^{(N)},\,\mathbf{x%
}_{l}^{(L)})=$ $\mathcal{N}(\mathbf{y}_{l};\mathbf{B}_{l}(\mathbf{x}%
_{l}^{(N)})\mathbf{x}_{l}^{(L)}+\mathbf{g}_{l}(\mathbf{x}_{l}^{(N)}),\mathbf{%
C}_{e})$ is given by%
\begin{equation}
\vec{m}_{ms}^{(k)}(\mathbf{x}_{l}^{(N)})=\int f(\mathbf{y}_{l}|\mathbf{x}%
_{l}^{(N)},\,\mathbf{x}_{l}^{(L)})\,\vec{m}_{fe}^{(k)}(\mathbf{x}%
_{l}^{(L)})\,d\mathbf{x}_{l}^{(L)}.  \label{eq:mess_ms_PF}
\end{equation}%
Based on CR2, it is easy to show that 
\begin{equation}
\vec{m}_{ms}^{(k)}(\mathbf{x}_{l}^{(N)})=\mathcal{N}\left( \mathbf{y}_{l};%
\mathbf{\tilde{\eta}}_{ms,l}^{(k)}\left( \mathbf{x}_{l}^{(N)}\right) ,%
\mathbf{\tilde{C}}_{ms,l}^{(k)}\left( \mathbf{x}_{l}^{(N)}\right) \right) ,
\label{eq:mess_ms_PF_expr}
\end{equation}%
where $\mathbf{\tilde{\eta}}_{ms,l}^{(k)}(\mathbf{x}_{l}^{(N)})\triangleq 
\mathbf{B}_{l}(\mathbf{x}_{l}^{(N)})\mathbf{\tilde{\eta}}_{fe2,l}^{(k)}+%
\mathbf{g}_{l}(\mathbf{x}_{l}^{(N)})$ and $\mathbf{\tilde{C}}_{ms,l}^{(k)}(%
\mathbf{x}_{l}^{(N)})\triangleq \mathbf{B}_{l}(\mathbf{x}_{l}^{(N)})\mathbf{%
\tilde{C}}_{fe2,l}^{(k)}\mathbf{B}_{l}^{T}(\mathbf{x}_{l}^{(N)})+\mathbf{C}%
_{e}$. Then, substituting (\ref{eq:message_fp_l_j_PF_bis}) and (\ref%
{eq:mess_ms_PF_expr})\ in (\ref{m_fe1_k_PF}) yields 
\begin{equation}
\vec{m}_{fe1,j}^{(k)}\left( \mathbf{x}_{l}^{(N)}\right)
=w_{fe1,l,j}^{(k)}\,\delta \left( \mathbf{x}_{l}^{(N)}-\mathbf{x}%
_{fp,l,j}^{(N)}[k-1]\right) ,  \label{eq:mess_fe_PF}
\end{equation}%
where\footnote{%
In evaluating the weight $w_{fe1,l,j}^{(k)}$ (\ref%
{eq:weight_before_resampling}), the factor $[\det (\mathbf{\tilde{C}}%
_{ms,l,j}^{(k)})]^{-P/2}$ appearing in the expression of the involved
Gaussian pdf \ is neglected in our simulations, since this entails a
negligible loss in estimation accuracy. Similar comments apply to the factor 
$\check{D}_{pm,l,j}^{(k)}$ appearing in the weight $w_{pm,l,j}^{(k)}$ (\ref%
{m_pm_x_N_l_j}).}
\begin{equation}
w_{fe1,l,j}^{(k)}\triangleq \mathcal{N}\left( \mathbf{y}_{l};\mathbf{\tilde{%
\eta}}_{ms,l,j}^{(k)},\mathbf{\tilde{C}}_{ms,l,j}^{(k)}\right)
\label{eq:weight_before_resampling}
\end{equation}%
is the new particle weight combining the a priori information about $\mathbf{%
x}_{l}^{(N)}$ with the information provided by the new measurement; here,%
\begin{equation}
\mathbf{\tilde{\eta}}_{ms,l,j}^{(k)}\triangleq \mathbf{\tilde{\eta}}%
_{ms,l}^{(k)}\left( \mathbf{x}_{fp,l,j}^{(N)}[k-1]\right) =\mathbf{B}%
_{l,j}[k]\,\mathbf{\tilde{\eta}}_{fe2,l}^{(k)}+\mathbf{g}_{l,j}
\label{eq:eta_fe_PF}
\end{equation}%
and%
\begin{eqnarray}
\mathbf{\tilde{C}}_{ms,l,j}^{(k)} &\triangleq &\mathbf{\tilde{C}}%
_{ms,l}^{(k)}\left( \mathbf{x}_{fp,l,j}^{(N)}[k-1]\right)  \notag \\
&=&\mathbf{B}_{l,j}[k]\mathbf{\tilde{C}}_{fe2,l}^{(k)}\left( \mathbf{B}%
_{l,j}[k]\right) ^{T}+\mathbf{C}_{e},  \label{eq:C_fe_PF}
\end{eqnarray}%
with $\mathbf{g}_{l,j}[k]\triangleq \mathbf{g}_{l}(\mathbf{x}%
_{fp,l,j}^{(N)}[k-1])$ and $\mathbf{B}_{l,j}[k]\triangleq \mathbf{B}_{l}(%
\mathbf{x}_{fp,l,j}^{(N)}[k-1])$.

3. \emph{Computation of the }PMs \emph{for} PF \emph{and} \emph{second }MU%
\emph{\ in} PF - This step aims at updating the weight of the $j$-th
particle $\mathbf{x}%
_{fp,l,j}^{(N)}[k-1]$ (provided by the message $\vec{m}_{fe1,j}^{(k)}(%
\mathbf{x}_{l}^{(N)})$ (\ref{eq:mess_fe_PF})) on the basis of the PM $%
\mathbf{z}_{l}^{(N)}$ (\ref{z_N_l}). It involves the computation of the PM
message $\vec{m}_{pm,j}^{(k)}(\mathbf{x}_{l}^{(N)})$ and of the message (see
Fig. \ref{Fig_5}) 
\begin{equation}
\vec{m}_{fe2,j}^{(k)}\left( \mathbf{x}_{l}^{(N)}\right) =\vec{m}%
_{fe1,j}\left( \mathbf{x}_{l}^{(N)}\right) \,\vec{m}_{pm,j}^{(k)}\left( 
\mathbf{x}_{l}^{(N)}\right) .  \label{m_fe2_k_PF}
\end{equation}%
The algorithm for computing $\vec{m}_{pm,j}^{(k)}(\mathbf{x}_{l}^{(N)})$ is
executed in the PMG$_{\text{EKF}}$ block shown in Figs. \ref{Fig_4}-\ref%
{Fig_5} and is described in detail in Appendix \ref{app:A}, where it is
shown that%
\begin{eqnarray}
&&w_{pm,l,j}^{(k)}\triangleq \vec{m}_{pm,j}^{(k)}\left( \mathbf{x}%
_{l}^{(N)}\right)  \notag \\
&=&\check{D}_{pm,l,j}^{(k)}\cdot \exp \left[ \frac{1}{2}\left( \left( 
\mathbf{\check{\eta}}_{pm,l,j}^{(k)}\right) ^{T}\mathbf{\check{W}}%
_{pm,l,j}^{(k)}\mathbf{\check{\eta}}_{pm,l,j}^{(k)}\right. \right.  \notag \\
&&\left. \left. -\left( \mathbf{\check{\eta}}_{z,l,j}^{(k)}\right) ^{T}%
\mathbf{\check{W}}_{z,l,j}^{(k)}\mathbf{\check{\eta}}_{z,l,j}^{(k)}-\left( 
\mathbf{f}_{l,j}^{(L)}[k]\right) ^{T}\mathbf{W}_{w}^{(L)}\mathbf{f}%
_{l,j}^{(L)}[k]\right) \right] ;  \label{m_pm_x_N_l_j}
\end{eqnarray}%
here%
\begin{equation}
\mathbf{\check{W}}_{pm,l,j}^{(k)}\triangleq \left( \mathbf{\check{C}}%
_{pm,l,j}^{(k)}\right) ^{-1}=\mathbf{\check{W}}_{z,l,j}^{(k)}+\mathbf{W}%
_{w}^{(L)},  \label{W_pm_x_N_l_j}
\end{equation}%
\begin{equation}
\mathbf{\check{w}}_{pm,l,j}^{(k)}\triangleq \mathbf{\check{W}}_{pm,l,j}^{(k)}%
\mathbf{\check{\eta}}_{pm,l,j}^{(k)}=\mathbf{\check{w}}_{z,l,j}^{(k)}+%
\mathbf{W}_{w}^{(L)}\mathbf{f}_{l,j}^{(L)},  \label{w_pm_x_N_l_j}
\end{equation}%
$\mathbf{\check{W}}_{z,l,j}^{(k)}\triangleq (\mathbf{\check{C}}%
_{z,l,j}^{(k)})^{-1}$, $\mathbf{\check{w}}_{z,l,j}^{(k)}\triangleq \mathbf{%
\check{W}}_{z,l,j}^{(k)}\mathbf{\check{\eta}}_{z,l,j}^{(k)}$ ($\mathbf{%
\check{\eta}}_{z,l,j}^{(k)}$ and $\mathbf{\check{C}}_{z,l,j}^{(k)}$ are
given by (\ref{eta_mess_z_N}) and (\ref{C_mess_Z_N_bis}), respectively), $%
\mathbf{W}_{w}^{(L)}\triangleq \lbrack \mathbf{C}_{w}^{(L)}]^{-1}$, $\mathbf{%
f}_{l,j}^{(L)}[k]\triangleq \mathbf{f}_{l}^{(L)}(\mathbf{x}%
_{fp,l,j}^{(N)}[k-1])$, $\check{D}_{pm,l,j}^{(k)}\triangleq \lbrack \det (%
\mathbf{\check{C}}_{l,j}^{(k)})]^{-D_{L}/2}$ and $\mathbf{\check{C}}%
_{l,j}^{(k)}\triangleq \mathbf{\check{C}}_{z,l,j}^{(k)}+\mathbf{C}_{w}^{(L)}$%
. Then, substituting (\ref{eq:mess_fe_PF}) and (\ref{m_pm_x_N_l_j}) in (\ref%
{m_fe2_k_PF}) yields 
\begin{equation}
\vec{m}_{fe2,j}^{(k)}\left( \mathbf{x}_{l}^{(N)}\right)
=w_{fe2,l,j}^{(k)}\,\delta \left( \mathbf{x}_{l}^{(N)}-\mathbf{x}%
_{fp,l,j}^{(N)}[k-1]\right) ,  \label{m_fe_2_x_N_l_bis}
\end{equation}%
where 
\begin{equation}
w_{fe2,l,j}^{(k)}\triangleq w_{fe1,l,j}^{(k)}\cdot w_{pm,l,j}^{(k)}
\label{w_fe_2_x_N_l}
\end{equation}%
represents the overall weight for the $j$-th particle of the set $%
S_{l}^{(k-1)}$; such a weight accounts for both the (real) measurement $%
\mathbf{y}_{l}$ and the PM $\mathbf{z}_{l}^{(N)}$ (through the weights $%
w_{fe1,l,j}$ and $w_{pm,l,j}^{(k)}$, respectively). Once all the weights $%
\{w_{fe2,l,j}^{(k)}\}$ are available, their normalization is accomplished;
this produces the normalised weights%
\begin{equation}
W_{fe2,l,j}^{(k)}\triangleq w_{fe2,l,j}^{(k)}\,K_{fe2,l}^{(k)},
\label{W_fe_2_x_N_l}
\end{equation}%
where $K_{fe2,l}^{(k)}\triangleq
1/\sum\limits_{l=0}^{N_{p}-1}w_{fe2,l,j}^{(k)}$. Note that the particles $\{%
\mathbf{x}_{fp,l,j}^{(N)}[k-1]\}$ and their new weights $\{W_{fe2,l,j}^{(k)}%
\}$ provide a statistical representation of the \emph{forward estimate} of $%
\mathbf{x}_{l}^{(N)}$ computed by PF\ in the $k$-th iteration.

Resampling with replacement is now accomplished for the particle set $%
S_{l}^{(k-1)}
$on the basis of the new weights $\{W_{fe2,l,j}^{(k)}\}$ (see (\ref%
{W_fe_2_x_N_l})). Note that this task does not emerge from the application
of SPA to the considered graphical model; however, it ensures that the
particles emerging from it are equally likely. Resampling simply entails
that the $N_{p}$ particles $\{\mathbf{x}_{fp,l,j}^{(N)}[k-1]\}$ and their
associated weights $\{W_{fe2,l,j}^{(k)}\}$ (\ref{W_fe_2_x_N_l}) are replaced
by the new particles $\{\mathbf{x}_{fp,l,j}^{(N)}[k]\}$, forming the\ set $%
S_{l}^{(k)}$ and having identical weights (all equal to $1/N_{p}$).
Consequently, the effect of resampling can be simply represented as turning
the message $\vec{m}_{fe2,j}^{(k)}(\mathbf{x}_{l}^{(N)})$ (\ref%
{m_fe_2_x_N_l_bis}) into 
\begin{equation}
\vec{m}_{fe2,j}^{(k)}\left( \mathbf{x}_{l}^{(N)}\right) =\delta \left( 
\mathbf{x}_{l}^{(N)}-\mathbf{x}_{fp,l,j}^{(N)}[k]\right) ,
\label{m_fe_2_x_N_l_tris}
\end{equation}%
with $j=0,1,...,N_{p}-1$.

4. \emph{Computation of the }PMs\emph{\ for} EKF - This step aims at
computing the Gaussian message 
\begin{equation}
\vec{m}_{pm}^{(k)}\left( \mathbf{x}_{l}\right) =\mathcal{\mathcal{N}}\left( 
\mathbf{x}_{l};\mathbf{\eta }_{pm,l}^{(k)},\mathbf{C}_{pm,l}^{(k)}\right) ,
\label{message_pm_x_l_k}
\end{equation}%
providing the PM information exploited by EKF\ in its second MU of the next
iteration. This requires combining the $N_{p}$ messages $\{\vec{m}%
_{fe2,j}^{(k)}(\mathbf{x}_{l}^{(N)})\}$ (see (\ref{m_fe_2_x_N_l_tris})) with
the $N_{p}$ \ messages $\{\vec{m}_{pm,j}^{(k)}(\mathbf{x}_{l}^{(L)})\}$,
evaluated in the PMG$_{\text{PF}}$ block appearing in Figs. \ref{Fig_4}-\ref%
{Fig_5} and conveying the (particle-dependent) statistical information
acquired about $\mathbf{x}_{l}^{(L)}$ on the basis of the PM $\mathbf{z}%
_{l}^{(L)}$ (\ref{eq:z_L_l}). The computation of the message $\vec{m}%
_{pm,j}^{(k)}(\mathbf{x}_{l}^{(L)})$ is described in detail in Appendix \ref%
{app:A}, where it is shown that%
\begin{equation}
\vec{m}_{pm,j}^{(k)}\left( \mathbf{x}_{l}^{(L)}\right) =\mathcal{\mathcal{N}}%
\left( \mathbf{x}_{l}^{(L)};\mathbf{\tilde{\eta}}_{pm,l,j}^{(k)},\mathbf{%
\tilde{C}}_{pm,l,j}^{(k)}\right) ;  \label{eq:message_pm_L_j_tris}
\end{equation}%
here, the covariance matrix $\mathbf{\tilde{C}}_{pm,l,j}^{(k)}$ and the mean
vector $\mathbf{\tilde{\eta}}_{pm,l,j}^{(k)}$ are computed on the basis of
the precision matrix%
\begin{equation}
\mathbf{\tilde{W}}_{pm,l,j}^{(k)}\triangleq \left( \mathbf{\tilde{C}}%
_{pm,l,j}^{(k)}\right) ^{-1}=\left( \mathbf{A}_{l,j}^{(N)}[k]\right) ^{T}%
\mathbf{W}_{w}^{(N)}\mathbf{A}_{l,j}^{(N)}[k]  \label{eq:W_pm_L_j}
\end{equation}%
and the transformed mean vector%
\begin{equation}
\mathbf{\tilde{w}}_{pm,l,j}^{(k)}\triangleq \mathbf{\tilde{W}}_{pm,l,j}^{(k)}%
\mathbf{\tilde{\eta}}_{pm,l,j}^{(k)}=\left( \mathbf{A}_{l,j}^{(N)}[k]\right)
^{T}\mathbf{W}_{w}^{(N)}\mathbf{z}_{l,j}^{(L)}[k],  \label{eq:w_pm_L_j}
\end{equation}%
respectively; moreover, $\mathbf{A}_{l,j}^{(N)}[k]\triangleq \mathbf{A}%
_{l}^{(N)}(\mathbf{x}_{fp,l,j}^{(N)}[k])$, $\mathbf{f}_{l,j}^{(N)}[k]%
\triangleq \mathbf{f}_{l}^{(N)}(\mathbf{x}_{fp,l,j}^{(N)}[k])$ and $\mathbf{z%
}_{l,j}^{(L)}[k]$ is defined by (\ref{eq:z_L_evaluated}).

The proposed technique for merging the information provided by $\{\vec{m}%
_{fe2,j}^{(k)}(\mathbf{x}_{l}^{(N)})\}$ (\ref{m_fe_2_x_N_l_tris}) with those
conveyed by $\{\vec{m}_{pm,j}^{(k)}(\mathbf{x}_{l}^{(L)})\}$ (\ref%
{eq:message_pm_L_j_tris}) is based on the following considerations. The
message $\vec{m}_{pm,j}^{(k)}(\mathbf{x}_{l}^{(L)})$ is \emph{coupled} with $%
\vec{m}_{fe2,j}^{(k)}(\mathbf{x}_{l}^{(N)})$ (for any $j$), since the
evaluation of the former message relies on the latter one (see Appendix \ref%
{app:A}). Moreover, these two messages provide \emph{complementary}
information, because they refer to the two different components of the
overall state $\mathbf{x}_{l}$. This explains why the joint statistical
information conveyed by the sets $\{\vec{m}_{fe2,j}^{(k)}(\mathbf{x}%
_{l}^{(N)})\}$ and $\{\vec{m}_{pm,j}^{(k)}(\mathbf{x}_{l}^{(L)})\}$ can be
expressed through the joint pdf%
\begin{equation}
f^{(k)}\left( \mathbf{x}_{l}^{(L)},\mathbf{x}_{l}^{(N)}\right) \triangleq 
\frac{1}{N_{p}}\sum\limits_{l=0}^{N_{p}-1}\vec{m}_{fe2,j}^{(k)}\left( 
\mathbf{x}_{l}^{(N)}\right) \vec{m}_{pm,j}^{(k)}\left( \mathbf{x}%
_{l}^{(L)}\right) .  \label{joint_pdf}
\end{equation}%
Then, the message $\vec{m}_{pm}^{(k)}(\mathbf{x}_{l})$ can be computed \emph{%
by projecting the last function onto a} \emph{single Gaussian pdf} (see (\ref%
{message_pm_x_l_k})), since message passing over the EKF portion of our
graphical model involves Gaussian messages only; the transformation adopted
here to achieve this result ensures that the \emph{mean }and the\emph{\
covariance} of the pdf $f^{(k)}(\mathbf{x}_{l}^{(L)},\mathbf{x}_{l}^{(N)})$ (%
\ref{joint_pdf}) are preserved\footnote{%
Details about the employed method for condensing the $N_{p}$-component \emph{%
Gaussian mixture} (GM) representing $\mathbf{x}_{l}^{(L)}$ into a single
Gaussian pdf can be found in \cite[Sec. IV]{Runnalls_2007}.}. For this
reason, if the mean $\mathbf{\eta }_{pm,l}^{(k)}\ $and the covariance matrix 
$\mathbf{C}_{pm,l}^{(k)}$ of the message $\vec{m}_{pm}^{(k)}\left( \mathbf{x}%
_{l}\right) $ (\ref{message_pm_x_l_k}) are put in the form 
\begin{equation}
\mathbf{\eta }_{pm,l}^{(k)}=\left[ \left( \mathbf{\tilde{\eta}}%
_{pm,l}^{(k)}\right) ^{T},\left( \mathbf{\check{\eta}}_{pm,l}^{(k)}\right)
^{T}\right] ^{T}  \label{eta_pm_l_k}
\end{equation}%
and%
\begin{equation}
\mathbf{C}_{pm,l}^{(k)}=\left[ 
\begin{array}{cc}
\mathbf{\tilde{C}}_{pm,l}^{(k)} & \mathbf{\dot{C}}_{pm,l}^{(k)} \\ 
\left( \mathbf{\dot{C}}_{pm,l}^{(k)}\right) ^{T} & \mathbf{\check{C}}%
_{pm,l}^{(k)}%
\end{array}%
\right]  \label{C_pm_l_k}
\end{equation}%
respectively, the $D_{L}$-dimensional mean vector $\mathbf{\tilde{\eta}}%
_{pm,l}^{(k)}$\ and the $D_{N}$-dimensional mean vector $\mathbf{\check{\eta}%
}_{pm,l}^{(k)}$ are computed as 
\begin{equation}
\mathbf{\tilde{\eta}}_{pm,l}^{(k)}\triangleq \frac{1}{N_{p}}%
\sum_{j=0}^{N_{p}-1}\mathbf{\tilde{\eta}}_{pm,l,j}^{(k)}
\label{eta_pm_l_L_k}
\end{equation}%
and%
\begin{equation}
\mathbf{\check{\eta}}_{pm,l}^{(k)}\triangleq \frac{1}{N_{p}}%
\sum_{j=0}^{N_{p}-1}\mathbf{x}_{fe,l,j}^{(N)}[k]  \label{eta_pm_l_N_k}
\end{equation}%
respectively, whereas the $D_{L}\times D_{L}$ \ covariance matrix $\mathbf{%
\tilde{C}}_{pm,l}^{(k)}$, the $D_{N}\times D_{N}$ \ covariance matrix $%
\mathbf{\check{C}}_{pm,l}^{(k)}$ and $D_{L}\times D_{N}$ \ cross-covariance
matrix $\mathbf{\dot{C}}_{pm,l}^{(k)}$ are computed as%
\begin{equation}
\mathbf{\tilde{C}}_{pm,l}^{(k)}\triangleq \frac{1}{N_{p}}\sum_{j=0}^{N_{p}-1}%
\mathbf{r}_{pm,l,j}^{(k)}-\mathbf{\tilde{\eta}}_{pm,l}^{(k)}\left( \mathbf{%
\tilde{\eta}}_{pm,l}^{(k)}\right) ^{T},  \label{C_pm_l_L_k_bis}
\end{equation}%
\begin{equation}
\mathbf{\check{C}}_{pm,l}^{(k)}\triangleq \frac{1}{N_{p}}\sum_{j=0}^{N_{p}-1}%
\mathbf{r}_{fe,l,j}^{(N)}[k]-\mathbf{\check{\eta}}_{pm,l}^{(k)}\left( 
\mathbf{\check{\eta}}_{pm,l}^{(k)}\right) ^{T},  \label{C_pm_l_N_k}
\end{equation}%
and%
\begin{equation}
\mathbf{\dot{C}}_{pm,l}^{(k)}\triangleq \frac{1}{N_{p}}\sum_{j=0}^{N_{p}-1}%
\mathbf{\dot{r}}_{pm,l,j}^{(k)}-\mathbf{\tilde{\eta}}_{pm,l}^{(k)}\left( 
\mathbf{\check{\eta}}_{pm,l}^{(k)}\right) ^{T},  \label{C_pm_l_LN_k}
\end{equation}%
respectively; here, $\mathbf{r}_{pm,l,j}^{(k)}\triangleq \mathbf{\tilde{C}}%
_{pm,l,j}^{(k)}+\mathbf{\tilde{\eta}}_{pm,l,j}^{(k)}(\mathbf{\tilde{\eta}}%
_{pm,l,j}^{(k)})^{T}$, $\mathbf{r}_{fe,l,j}^{(N)}[k]\triangleq \mathbf{x}%
_{fe,l,j}^{(N)}[k](\mathbf{x}_{fe,l,j}^{(N)}[k])^{T}$ and $\mathbf{\dot{r}}%
_{pm,l,j}^{(k)}\triangleq \mathbf{\tilde{\eta}}_{pm,l,j}^{(k)}(\mathbf{x}%
_{fe,l,j}^{(N)}[k])^{T}$. The evaluation of the parameters $\mathbf{\eta }%
_{pm,l}^{(k)}$ (\ref{eta_pm_l_k}) and $\mathbf{C}_{pm,l}^{(k)}$ (\ref%
{C_pm_l_k}) for the message $\vec{m}_{pm}^{(k)}\left( \mathbf{x}_{l}\right) $
(\ref{message_pm_x_l_k}) concludes step 4. (i.e., the last step of the $k$%
-th iteration). This message is stored for the next iteration; then, if the
iteration index $k$ is less than $N_{it}$, it is increased by one, so that a
new iteration can be started by going back to step $1$. On the contrary, if $%
k=N_{it}$, the message (see (\ref{m_fe2_EKF_k})-(\ref{w_fe2_EKF_k}) and Fig. %
\ref{Fig_5}) 
\begin{equation}
\vec{m}_{fe2,l}^{(N_{it}+1)}\left( \mathbf{x}_{l}\right) =\mathcal{\mathcal{N%
}}\left( \mathbf{x}_{l};\mathbf{\eta }_{fe2,l}^{(N_{it}+1)},\mathbf{C}%
_{fe2,l}^{(N_{it}+1)}\right) ,  \label{m_fe2_EKF_final}
\end{equation}%
is computed as if a new iteration was started. Finally, if $l<t$, the output
messages $\{\vec{m}_{fp,j}(\mathbf{x}_{l+1}^{(N)})\}$ and $\vec{m}%
_{fp}\left( \mathbf{x}_{l+1}\right) $ (i.e., the new predictions of the two
state components) are computed. On the one hand, the message $\vec{m}_{fp,j}(%
\mathbf{x}_{l+1}^{(N)})$ is easily generated as (see (\ref{eq:m_fp_N_l+1_bis}%
)-(\ref{eq:C_fp_N_l+1})) 
\begin{equation}
\vec{m}_{fp,j}\left( \mathbf{x}_{l+1}^{(N)}\right) =\vec{m}%
_{fp,j}^{(N_{it})}\left( \mathbf{x}_{l+1}^{(N)}\right)
\label{m_fp_N_l+1_final}
\end{equation}%
for $j=0,1,...,N_{p}-1$. On the other hand, $\vec{m}_{fp}\left( \mathbf{x}%
_{l+1}\right) $ is computed as (see Fig. \ref{Fig_5}) 
\begin{equation}
\vec{m}_{fp}\left( \mathbf{x}_{l+1}\right) =\int \,\tilde{f}\left( \mathbf{x}%
_{l+1}\left\vert \mathbf{x}_{l}\right. \right) \,\vec{m}%
_{fe2,l}^{(N_{it}+1)}\left( \mathbf{x}_{l}\right) \,d\mathbf{x}_{l}.
\label{m_fp_l+1_whole}
\end{equation}%
Since $\tilde{f}\left( \mathbf{x}_{l+1}\left\vert \mathbf{x}_{l}\right.
\right) =\mathcal{N}(\mathbf{x}_{l+1};\mathbf{F}_{l}\mathbf{x}_{l}+\mathbf{u}%
_{l},\mathbf{C}_{w}\mathcal{)}$ (see (\ref{state_up_approx})) and $\vec{m}%
_{fe2,l}^{(N_{it}+1)}(\mathbf{x}_{l})$ is a Gaussian message (see (\ref%
{m_fe2_EKF_final})), applying CR2 to the evaluation of the RHS\ of (\ref%
{m_fp_l+1_whole}) produces 
\begin{equation}
\vec{m}_{fp}\left( \mathbf{x}_{l+1}\right) =\mathcal{N}\left( \mathbf{x}%
_{l+1};\mathbf{\eta }_{fp,l+1},\mathbf{C}_{fp,l+1}\right) \,,
\label{m_fp_l+1_whole_bis}
\end{equation}%
where%
\begin{equation}
\mathbf{\eta }_{fp,l+1}\triangleq \mathbf{F}_{l}\,\mathbf{\eta }%
_{fe2,l}^{(N_{it}+1)}+\mathbf{u}_{l}  \label{eta_fp_l+1_whole}
\end{equation}%
and%
\begin{equation}
\mathbf{C}_{fp,l+1}\triangleq \mathbf{C}_{w}+\mathbf{F}_{l}\mathbf{C}%
_{fe2,l}^{(N_{it}+1)}\mathbf{F}_{l}^{T}.  \label{C_fp_l+1_whole}
\end{equation}%
The $l$-th recursion is now over.

The algorithm described above needs a proper initialization. In our work,
the Gaussian pdf $f(\mathbf{x}_{1})=\mathcal{\mathcal{N(}}\mathbf{x}_{1};%
\mathbf{\eta }_{1},\mathbf{C}_{1})$ is assumed for $\mathbf{x}_{1}$.
Consequently, as far as PF is concerned, before starting the first recursion
(corresponding to $l=1$), the set $S_{1}=\{\mathbf{x}_{fp,1,j}^{(N)},%
\,j=0,1,...,N_{p}-1\}$ is generated for $\mathbf{x}_{1}^{(N)}$ by sampling
the pdf $f(\mathbf{x}_{1}^{(N)})$ (that results from the marginalization of $%
f(\mathbf{x}_{1})$ with respect to $\mathbf{x}_{1}^{(L)}$) $N_{p}$ times;
then, the same weight is assigned to each particle (i.e., $%
w_{fp,1,j}=1/N_{p} $ for any $j$). Moreover, we set $\vec{m}_{fp}\left( 
\mathbf{x}_{1}\right) =f(\mathbf{x}_{1})$ for the EKF portion of the TF\#1
algorithm.

All the processing tasks accomplished in the message passing procedure
derived above are summarized in Algorithm 1. Note also that, at the end of
the $l$-th recursion, estimates of $\mathbf{x}_{l}^{(N)}$ and $\mathbf{x}%
_{l}^{(L)}$ can be evaluated as: a) $\mathbf{\hat{x}}_{l}^{(N)}=%
\sum_{j=0}^{N_{p}-1}W_{fe2,l,j}^{(N_{it})}\mathbf{x}%
_{fp,l,j}^{(N)}[N_{it}-1] $ (see our previous comments following eq. (\ref%
{W_fe_2_x_N_l})) or $\mathbf{\hat{x}}_{l}^{(N)}=\mathbf{\bar{\eta}}%
_{fe2,l}^{(N_{it}+1)}$, where $\mathbf{\bar{\eta}}_{fe2,l}^{(N_{it}+1)}$
consists of the last $D_{N}$ elements of $\mathbf{\eta }%
_{fe2,l}^{(N_{it}+1)} $ (see (\ref{m_fe2_EKF_final})); b) $\mathbf{\hat{x}}%
_{l}^{(L)}=\mathbf{\hat{\eta}}_{fe2,l}^{(N_{it}+1)}$, where $\mathbf{\tilde{%
\eta}}_{fe2,l}^{(N_{it}+1)}$ consists of the first $D_{L}$ elements of $%
\mathbf{\eta }_{fe2,l}^{(N_{it}+1)}$.

\begin{algorithm}
	\SetKw{a}{a-}
	\SetKw{b}{b-}
	\SetKw{c}{c-}
	\SetKw{d}{d-}
	\SetKw{e}{d1-}
	\SetKw{f}{d2-}
	\SetKw{g}{d3-}
	\SetKw{h}{e-}
	\SetKw{i}{f-}
	\SetKw{j}{g-}
	\SetKw{k}{h-}
		
	\nl\textbf{Initialisation:} 
	For $j=0$ to $N_{p}-1$: sample the pdf $f(\mathbf{x}_{1}^{(N)})$ to generate the particles $\mathbf{x}_{fp,1,j}^{(N)}$ (forming $S_1^{(0)}$), and
	assign the weight $w_{fp,1}=1/N_{p}$ to each of them. Set $\mathbf{W}_{fp,1}=\mathbf{W}_{1}=[\mathbf{C}_{1}]^{-1}$, $\mathbf{w}_{fp,1}=\mathbf{W}_{1}{\eta}_{1}$.
	
	\nl\textbf{Filtering:} For $l=1$ to $t$:
	
	\a \emph{First} MU in EKF: Compute $\mathbf{W}_{fe1,l}$ (\ref{W_fe1_EKF}) and $\mathbf{w}_{fe1,l}$ (\ref{w_fe1_EKF}), $\mathbf{C}_{fe1,l}=[\mathbf{W}_{fe1,l}]^{-1}$ and $\mathbf{\eta }_{fe1,l}=\mathbf{C}_{fe1,l}\mathbf{w}_{fe1,l}$. Then, extract $\mathbf{\tilde{\eta}}_{fe1,l}$ and $\mathbf{\tilde{C}}_{fe1,l}$ from $\mathbf{\eta}_{fe1,l}$ and $\mathbf{C}_{fe1,l}$, respectively. Set $\mathbf{W}_{pm,l}^{(0)}=\mathbf{0}_{D,D}$ and $\mathbf{w}_{pm,l}^{(0)}=\mathbf{0}_{D}$.
	
	\For {$k=1$ to $N_{it}$}{
		
		\b  \emph{Second} MU in EKF. Compute $\mathbf{C}_{fe2,l}^{(k)}$ (\ref{W_fe2_EKF_k}) and $\mathbf{\eta}_{fe2,l}^{(k)}$ (\ref{w_fe2_EKF_k}).
		
		\c \emph{Marginalization}: extract $\mathbf{\tilde{\eta}}_{fe2,l}^{(k)}$ and $\mathbf{\tilde{C}}_{fe2,l}^{(k)}$ from $\mathbf{\eta}_{fe2,l}^{(k)}$ and $\mathbf{C}_{fe2,l}^{(k)}$, respectively.
		
		\d MUs in PF:\\
		\For {$j=0$ to $N_{p}-1$} 
		{\e \emph{First} MU in PF: compute $\mathbf{\tilde{\eta}}_{ms,l,j}^{(k)}$ (\ref{eq:eta_fe_PF}), $\mathbf{\tilde{C}}_{ms,l,j}^{(k)}$ (\ref{eq:C_fe_PF}) and $w_{fe1,l,j}^{(k)}$ (\ref{eq:weight_before_resampling}).

		\f  \emph{Computation of} PMs for PF: compute $\mathbf{\check{\eta}}_{z,l,j}^{(k)}$ (\ref{eta_mess_z_N}) and $\mathbf{\check{C}}_{z,l,j}^{(k)}$ (\ref{C_mess_Z_N_bis}), $\mathbf{\check{W}}_{z,l,j}^{(k)}= [\mathbf{\check{C}}%
		_{z,l,j}^{(k)}]^{-1}$ and $\mathbf{\check{w}}_{z,l,j}^{(k)}= \mathbf{\check{W}}_{z,l,j}^{(k)}\mathbf{\check{\eta}}_{z,l,j}^{(k)}$. Then, compute $\mathbf{\check{W}}_{pm,l,j}^{(k)}$ (\ref{W_pm_x_N_l_j}) and $\mathbf{\check{w}}_{pm,l,j}^{(k)}$ (\ref{w_pm_x_N_l_j}), $\mathbf{\check{C}}_{pm,l,j}^{(k)} = [\mathbf{\check{W}}_{pm,l,j}^{(k)}]^{-1}$ and ${\check{\eta}}_{pm,l,j}^{(k)}=\mathbf{\check{C}}_{pm,l,j}^{(k)} \mathbf{\check{w}}_{pm,l,j}^{(k)}$. Finally, compute $ w_{pm,l,j}^{(k)}$ (\ref{m_pm_x_N_l_j}).
			
		\g \emph{Second} MU in PF: compute $w_{fe2,l,j}^{(k)}$ (\ref{w_fe_2_x_N_l}). 	
	}
		\h \emph{Normalization of particle weights}: compute the normalized weights $\{W_{fe2,l,j}^{(k)}\}$ according to (\ref{W_fe_2_x_N_l}).
			
		\i \emph{Resampling with replacement}: generate the new particle set $S_l^{(k)} = \{\mathbf{x}_{fp,l,j}^{(N)}[k]\}$ by resampling $S_l^{(k-1)} $ on the basis of the weights $\{W_{fe2,l,j}^{(k)}\}$.
			
		\j  \emph{Computation of} PM for EKF: For $j=1$ to $N_{p}$: Compute $\mathbf{\check{\eta}}_{fp,l,j}^{(k)}$ (\ref{eq:eta_fp_N_l+1}) and $\mathbf{\check{C}}_{fp,l,j}^{(k)}$ (\ref{eq:C_fp_N_l+1}), and sample the pdf $\mathcal{N}( 
		\mathbf{x}_{l+1}^{(N)};\mathbf{\check{\eta}}_{fp,l,j}^{(k)},\mathbf{\check{C}%
		}_{fp,l,j}^{(k)})$ to generate the new particle $\mathbf{x}_{fp,l+1,j}^{(N)}[k]$ and assign the weight $1/N_{p}$ to it.
		Then, compute $\mathbf{z}_{l,j}^{(L)}\left[ k\right]$ (\ref{eq:z_L_evaluated}), $\mathbf{\tilde{W}}_{pm,l,j}^{(k)} $ (\ref{eq:W_pm_L_j}) and $\mathbf{\tilde{w}}_{pm,l,j}^{(k)} $ (\ref{eq:w_pm_L_j}), $\mathbf{\tilde{C}}_{pm,l,j}^{(k)}=[\mathbf{\tilde{W}}_{pm,l,j}^{(k)}]^{-1}$ and ${\tilde{\eta}}_{pm,l,j}^{(k)} =\mathbf{\tilde{C}}_{pm,l,j}^{(k)} \mathbf{\tilde{w}}_{pm,l,j}^{(k)}$. Finally, compute $\mathbf{\eta }_{pm,l}^{(k)} $ (\ref{eta_pm_l_k}) and $\mathbf{C}_{pm,l}^{(k)} $ (\ref{C_pm_l_k}) (according to (\ref{eta_pm_l_L_k})-(\ref{C_pm_l_LN_k})), $\mathbf{W}_{pm,l}^{(k)}=[\mathbf{C}_{pm,l}^{(k)}]^{-1} $ and $\mathbf{w}_{pm,l}^{(k)}=\mathbf{W}_{pm,l}^{(k)} {\eta}_{pm,l}^{(k)}$ .	
		}
		
		\k \emph{Compute forward prediction} (if $l<t$): For $j=1$ to $N_{p}$: set $\mathbf{x}_{fp,l+1,j}^{(N)}=\mathbf{x}_{fp,l+1,j}^{(N)}[N_{it}]$ (these particles form the set $S_{l+1}$) and the weight $W_{fe2,l+1,j}=W_{fe2,l+1,j}^{(N_{it})}$. %
		Compute  $\mathbf{C}_{fe2,l}^{(N_{it}+1)}$ %
		and $\mathbf{\eta}_{fe2,l}^{(N_{it}+1)}$ on the basis of (\ref{W_fe2_EKF_k}) and (\ref{w_fe2_EKF_k}). Then, compute $\mathbf{\eta }_{fp,l+1}$ (\ref{eta_fp_l+1_whole}) and $\mathbf{C}_{fp,l+1}$ (\ref{C_fp_l+1_whole}), $\mathbf{W}_{fp,l+1}=[\mathbf{C}_{fp,l+1}]^{-1}$ and $\mathbf{w}_{fp,l+1}=$ $\mathbf{W}_{fp,l+1} {\eta}_{fp,l+1}$.

		\caption{Turbo Filtering \#1}
		\label{alg:TF1}
	\end{algorithm}

\begin{algorithm}
		\SetKw{a}{a-}
		\SetKw{b}{b-}
		\SetKw{c}{b1-}
		\SetKw{d}{b2-}
		\SetKw{e}{c-}
		\SetKw{f}{d-}
		\SetKw{g}{e-}
		\SetKw{h}{f-}
		\SetKw{i}{g-}
		\SetKw{j}{h-}
		\SetKw{k}{i-}
		
		\nl\textbf{Initialisation:} 
		Same as Alg. \ref{alg:TF1}.

		\nl\textbf{Filtering:} For $l=1$ to $t$:
		
		\a \emph{First} MU in EKF: Same as Alg. \ref{alg:TF1}, task \textbf{a}.
		
		\For {$k=1$ to $N_{it}$}{
			
			\b MUs in PF:\\
			\For {$j=0$ to $N_{p}-1$} 
			{\c \emph{First} MU in PF: Same as Alg. \ref{alg:TF1}, task \textbf{d1}.
				
			\d \emph{Second} MU in PF: Same as Alg. \ref{alg:TF1}, task \textbf{d3}.	
			}
			\e \emph{Normalization of particle weights}: Same as Alg. \ref{alg:TF1}, task \textbf{e}.
			
			\f \emph{Resampling with replacement}: Same as Alg. \ref{alg:TF1}, task \textbf{f}.
			
			\g  \emph{Computation of} PM for EKF: Same as Alg. \ref{alg:TF1}, task \textbf{g}.

			\h  \emph{Second} MU in EKF: Same as Alg. \ref{alg:TF1}, task \textbf{b}.
			
			\i \emph{Marginalization}: Same as Alg. \ref{alg:TF1}, task \textbf{c}.
			
			\j  \emph{Computation of} PMs for PF: Same as Alg. \ref{alg:TF1}, task \textbf{d2}.				
		}
		
		\k \emph{Compute forward prediction} (if $l<t$): 
		
		\For {$j=0$ to $N_{p}-1$}{
		Compute $\mathbf{\tilde{\eta}}_{ms,l,j}^{(k)}$ (\ref{eq:eta_fe_PF}), $\mathbf{\tilde{C}}_{ms,l,j}^{(k)}$ (\ref{eq:C_fe_PF}) and $w_{fe1,l,j}^{(k)}$ (\ref{eq:weight_before_resampling}), than compute $w_{fe2,l,j}^{(k)}$ (\ref{w_fe_2_x_N_l}).
			}	
		Finally, compute $\mathbf{\eta }_{fp,l+1}$ (\ref{eta_fp_l+1_whole}) and $\mathbf{C}_{fp,l+1}$ (\ref{C_fp_l+1_whole}), $\mathbf{W}_{fp,l+1}=[\mathbf{C}_{fp,l+1}]^{-1}$ and $\mathbf{w}_{fp,l+1}=$ $\mathbf{W}_{fp,l+1} {\eta}_{fp,l+1}$.
		
		\caption{Turbo Filtering \#2}
	\end{algorithm}

The scheduling adopted in the $k$-th iteration of the $l$-the recursion
accomplished by TF\#2 consists in computing the involved messages according
to the following order: 1) $\{\vec{m}_{ms,j}^{(k)}(\mathbf{x}_{l}^{(N)})\}$, 
$\{\vec{m}_{fe1,j}^{(k)}(\mathbf{x}_{l}^{(N)})\}$ (first MU in PF); 2) $\{%
\vec{m}_{fe2,j}^{(k)}(\mathbf{x}_{l}^{(N)})\}$ (second MU in PF; note that $%
\vec{m}_{pm,j}^{(0)}(\mathbf{x}_{l}^{(N)})=1$ for any $j$); 3) $\{\vec{m}%
_{pm,j}^{(k)}(\mathbf{x}_{l}^{(L)})\}$, $\vec{m}_{pm}^{(k)}(\mathbf{x}_{l})$%
, $\vec{m}_{fe2}^{(k)}\left( \mathbf{x}_{l}\right) $, $\vec{m}_{fe2}^{(k)}(%
\mathbf{x}_{l}^{(L)})$\thinspace\ (computation of PMs for EKF and second MU
in EKF); 4) $\{\vec{m}_{pm,j}^{(k)}(\mathbf{x}_{l}^{(N)})\}$ (computation of
PMs for PF). This algorithm can be easily derived following the same line of
reasoning as TF\#1 and is summarised in Algorithm 2.

As far as the computational complexity of TF\#1 and TF\#2 is concerned, it
can be shown that it is of order $\mathcal{O}(N_{TF})$, with%
\begin{eqnarray}
N_{TF} &=&2DP^{2}+PD^{2}+(N_{it}+4)D^{3}  \notag \\
&&+N_{it}\cdot N_{p}(PD_{L}^{2}+P^{2}D_{L}+P^{3}  \notag \\
&&+6D_{L}^{3}+2D_{N}D_{L}^{2}+3D_{L}D_{N}^{2}+D_{N}^{3}/3).
\label{TF_comp_compl}
\end{eqnarray}%
The last expression has been derived keeping into account all the dominant
contributions due to matrix inversions, matrix products and Cholesky
decompositions, that need to be accomplished for the complete state update
and measurement models expressed by (\ref{eq:XL_update}) and (\ref{eq:y_t}),
respectively. However, all the possible contributions originating from the
evaluation of the matrices $\mathbf{A}_{l}^{(Z)}(\mathbf{x}_{l}^{(N)})$ and
the functions $\mathbf{f}_{l}^{(Z)}(\mathbf{x}_{l}^{(N)})$ (with $Z=L$ and $%
N $) over the considered particle sets are not accounted for. A similar
approach has been followed for MPF, whose complexity\footnote{%
An assessment of MPF complexity is also available in \cite%
{Schon_2005_complexity}.} is of order $\mathcal{O}(N_{MPF})$, with 
\begin{eqnarray}
N_{MPF} &=&N_{p}(2PD_{L}^{2}+3P^{2}D_{L}+P^{3}+5D_{L}^{3}  \notag \\
&&+2D_{L}^{2}D_{N}+3D_{L}D_{N}^{2}+D_{N}^{3}/3).  \label{MPF_comp_compl}
\end{eqnarray}

Finally, it is worth mentioning that TF\#1 and TF\#2 have substantially
smaller memory requirements than MPF; in fact, the former algorithms need to
store the state estimates generated by a single extended Kalman filter,
whereas the latter one those computed by $N_{p}$ Kalman filters running in
parallel. This means that, if MPF is employed, a larger number of memory
accesses must be accomplished on the hardware platform on which the
filtering algorithm is run; as evidenced by our numerical results, this
feature can make the overall execution time of MPF much larger than that
required by TF, even if $N_{TF}>N_{MPF}$ for the same value of $N_{p}$.

\section{Interpretation of Turbo Filtering\label{Interpretation}}

An interesting interpretation of the processing tasks accomplished by the
TF\#1 and TF\#2 algorithms can be developed as follows. In TF\#1, the $j$-th
particle weight $w_{fe2,l,j}^{(k)}$ (\ref{w_fe_2_x_N_l}) available at the
end of the second MU of PF expresses the \emph{a posteriori} statistical
information about the particle $\mathbf{x}_{fp,l,j}^{(N)}[k-1]$ and can be
put in the equivalent form 
\begin{equation}
w_{fe2,l,j}^{(k)}=w_{l,j}^{(a)}\cdot w_{fe1,l,j}^{(k)}\cdot w_{pm,l,j}^{(k)}%
\text{;}  \label{m_4}
\end{equation}%
where $w_{l,j}^{(a)}$ denotes the \emph{a priori} information available for
the particle itself (in our derivation $w_{l,j}^{(a)}$ $=1$ has been
assumed, in place of $w_{l,j}^{(a)}$ $=1/N_{p}$, to simplify the notation;
see (\ref{eq:message_fp_l_j_PF})). Taking the natural logarithm of both
sides of (\ref{m_4}) produces 
\begin{equation}
L_{l,j}[k]=L_{l,j}^{(a)}+L_{l,j}^{(y)}[k]+L_{l,j}^{(z)}[k]  \label{loglik}
\end{equation}%
where $L_{l,j}[k]\triangleq \ln (w_{fe2,l,j}^{(k)})$, $L_{l,j}^{(a)}%
\triangleq \ln (w_{l,j}^{(a)})$, $L_{l,j}^{(y)}[k]\triangleq \ln
(w_{fe1,l,j}^{(k)})$ and $L_{l,j}^{(z)}\triangleq \ln (w_{pm,l,j}^{(k)})$.
The last equation has the same mathematical structure as the well known
formula (see \cite[Sec. 10.5, p. 450, eq. (10.15)]{Vitetta} or \cite[Par.
II.C, p. 432, eq. (20)]{Hagenauer_1996}) 
\begin{equation}
L\left( u_{j}|\mathbf{y}\right) =L\left( u_{j}\right)
+L_{c}(y_{j})+L_{e}\left( u_{j}\right)   \label{log_turbo}
\end{equation}%
expressing of the \emph{log-likelihood ratio} (LLR) available for the $j$-th
information bit\ $u_{j}$ at the output of a SISO channel decoder operating\ over an \emph{additive white Gaussian noise}
(AWGN) channel and fed by: a) the channel output vector $\mathbf{y}$ (whose $%
j$-th element $y_{j}$ is generated by the communications channel in response
to a channel symbol conveying $u_{j}$ and is processed to produce the
so-called \emph{channel} LLR $L_{c}(y_{j})$); b) the a priori LLR $L\left(
u_{j}\right) $ about $u_{j}$; c) the \emph{extrinsic} LLR $L_{e}\left(
u_{j}\right) $, i.e. a form of soft information available about $u_{j}$, but
intrinsically not influenced by such a bit (in turbo decoding of
concatenated channel codes extrinsic infomation is generated by another
channel decoder with which soft information is exchanged with the aim of
progressively refining data estimates). This correspondence is not only
formal, since the term $L_{l,j}^{(y)}[k]$ ($L_{l,j}^{(a)}$) in (\ref{loglik}%
) provides the same kind of information as $L_{c}(y_{j})$ ($L\left(
u_{j}\right) $), since these are both related to the noisy data (a priori
information) available about the quantities to be estimated (the system
state in one case, an information bit in the other one). These
considerations suggest that the term $L_{l,j}^{(z)}[k]$ of (\ref{loglik})
should represent the counterpart of the quantity $L_{e}\left( u_{j}\right) $
appearing in (\ref{log_turbo}), i.e. the so called \emph{extrinsic
information} (in other words, that part of the information\ available about $%
\mathbf{x}_{l}^{(N)}$ and not \emph{intrinsically} influenced by $\mathbf{x}%
_{l}^{(N)}$ itself). This interpretation is confirmed by the fact that $%
L_{l,j}^{(z)}[k]$ is computed on the basis of the statistical knowledge
available about $\mathbf{x}_{l}^{(L)}$ and $\mathbf{x}_{l+1}^{(L)}$ (see
Appendix \ref{app:A}), which, thanks to (\ref{eq:XL_update}) (with $Z=L$),
does provide useful information about $\mathbf{x}_{l}^{(N)}$. \ 

The reader can easily verify that an interpretation similar to that provided
for $w_{fe2,l,j}^{(k)}$ (\ref{w_fe_2_x_N_l}) can be given for $\vec{m}%
_{fe2}^{(k)}(\mathbf{x}_{l})$ (\ref{m_fe2_EKF_k}) (that conveys our \emph{a
posteriori} information about $\mathbf{x}_{l}$). In fact, the last message
results from the product of the messages $\vec{m}_{fp}\left( \mathbf{x}%
_{l}\right) $ (\ref{eq:eq:message_fp_l_EKF}), $\vec{m}_{fe1}^{(k)}(\mathbf{x}%
_{l})$ (\ref{m_fe1_EKF}) and $\vec{m}_{pm}^{(k)}\left( \mathbf{x}_{l}\right) 
$ (\ref{message_pm_x_l_k}); these convey \emph{prior}, \emph{measurement}
and \emph{extrinsic} information about $\mathbf{x}_{l}$, respectively. It is
worth noting, however, that $\vec{m}_{pm}^{(k)}\left( \mathbf{x}_{l}\right) $
(\ref{message_pm_x_l_k}) combines two different contributions, namely the
contributions from the message sets $\{\vec{m}_{fe2,j}^{(k)}(\mathbf{x}%
_{l}^{(N)})\}$ (\ref{m_fe_2_x_N_l_tris}) and $\{\vec{m}_{pm,j}^{(k)}(\mathbf{%
x}_{l}^{(L)})\}$ (\ref{eq:message_pm_L_j_tris}); however, only the message $%
\vec{m}_{pm,j}^{(k)}(\mathbf{x}_{l}^{(L)})$ can be really interpreted as the
counterpart of $w_{pm,l,j}^{(k)}$ (\ref{m_pm_x_N_l_j}), since its
computation is based on the PM message $\vec{m}_{j}^{(k)}(\mathbf{z}%
_{l}^{(L)})$ (\ref{m_j_z_L_bis}). 

\section{Numerical Results\label{num_results}}

In this Section we compare, in terms of accuracy and execution time, the
TF\#1 and TF\#2 algorithms with EKF and MPF for a specific CLG SSM. The
considered SSM refers to an agent moving on a plane and whose state $\mathbf{%
x}_{l}$ in the $l$-th observation interval is defined as $\mathbf{x}%
_{l}\triangleq \lbrack \mathbf{p}_{l}^{T},\mathbf{v}_{l}^{T}]^{T}$, where $%
\mathbf{v}_{l}\triangleq \lbrack v_{x,l},v_{y,l}]^{T}$ and $\mathbf{p}%
_{l}\triangleq \lbrack p_{x,l},p_{y,l}]^{T}$ represent the agent velocity
and its position, respectively (their components are expressed in m/s and in
m, respectively). As far as the state update equations are concerned, we
assume that: a) the agent velocity is approximately constant within each
sampling interval; b) the model describing its time evolution is obtained by
including the contribution of a \emph{position- and velocity-dependent force}
in a first-order autoregressive model (characterized by the \emph{forgetting
factor} $\rho $, with $0<\rho <1$). Therefore, the dynamic model 
\begin{equation}
\mathbf{v}_{l+1}=\rho \mathbf{v}_{l}+\left( 1-\rho \right) \mathbf{n}%
_{v,l}+\,\mathbf{a}_{l}\left( \mathbf{p}_{l},\mathbf{v}_{l}\right) T_{s},
\label{mod_1_v}
\end{equation}%
is adopted for velocity; here, $\{\mathbf{n}_{v,l}\}$ is an \emph{additive
white Gaussian noise} (AWGN) process (whose elements are characterized by
the covariance matrix $\mathbf{I}_{2}$), $T_{s}$ is the sampling interval and%
\begin{equation}
\mathbf{a}_{l}\left( \mathbf{p}_{l},\mathbf{v}_{l}\right) =-(a_{0}/d_{0})%
\mathbf{p}_{l}-\tilde{a}_{0}f_{v}\left( \left\Vert \mathbf{v}_{l}\right\Vert
\right) \mathbf{u}_{v,l}.  \label{acc_SSM1}
\end{equation}%
In the RHS\ of the last formula, $a_{0}$ and $\tilde{a}_{0}$ are scale
factors (both expressed in m/s$^{2}$), $d_{0}$ is a \emph{reference
distance, }$\mathbf{u}_{v,l}\triangleq \mathbf{v}_{l}/\left\Vert \mathbf{v}%
_{l}\right\Vert $ is the versor associated with $\mathbf{v}_{l}$ and $%
f_{v}\left( x\right) =(x/v_{0})^{3}$ is a continuous, differentiable and
dimensionless function expressing the dependence of the second term on the
intensity of $\mathbf{v}_{l}$ (the parameter $v_{0}$ represents a \emph{%
reference velocity}). Note that the first term and the second one in the
RHS\ of (\ref{acc_SSM1}) represent the contribution of \emph{%
position-dependent force} pointing towards the origin and proportional to $%
\left\Vert \mathbf{p}_{l}\right\Vert $, and that of \emph{velocity-dependent
force }acting as a resistance to the motion of the agent, respectively.

Given (\ref{mod_1_v}), the dynamic model 
\begin{equation}
\mathbf{p}_{l+1}=\mathbf{p}_{l}+\mathbf{v}_{l}T_{s}+\frac{1}{2}\mathbf{a}%
_{l}\left( \mathbf{p}_{l},\mathbf{v}_{l}\right) T_{s}^{2}+\mathbf{n}_{p,l}
\label{mod_1_p}
\end{equation}%
can be employed for the position of the considered agent; here, $\{\mathbf{n}%
_{p,l}\}$ is an AWGN process (whose elements are characterized by the
covariance matrix $\sigma _{p}^{2}\mathbf{I}_{2}$), independent if $\{%
\mathbf{n}_{v,l}\}$ and accounting for model inaccuracy.

In our study the measurement model 
\begin{equation}
\mathbf{y}_{l}=[\mathbf{p}_{l}^{T}\,\left\Vert \mathbf{v}_{l}\right\Vert
]^{T}+\mathbf{e}_{l},  \label{mod_1_y}
\end{equation}%
is also adopted; here, $\{\mathbf{e}_{l}\}$ is an AWGN process, whose
elements are characterized by the covariance matrix $\mathbf{C}_{e}=$diag$%
(\sigma _{e,p}^{2},\sigma _{e,p}^{2},\sigma _{e,v}^{2})$. \ Then, if we set $%
\mathbf{x}_{l}^{(L)}=\mathbf{p}_{l}$ and $\mathbf{x}_{l}^{(N)}=\mathbf{v}%
_{l} $, it is not difficult to show that the state equation (\ref{mod_1_v})
((\ref{mod_1_p})) and the measurement equation (\ref{mod_1_y}) can been
considered as instances of (\ref{eq:XL_update}) with $Z=L$ ((\ref%
{eq:XL_update}) with $Z=N$) and (\ref{eq:y_t}), respectively.

In our computer simulations, the estimation accuracy of the considered
filtering techniques has been assessed by evaluating two \emph{root mean
square errors} (RMSEs), one for the linear state component, the other for
the nonlinear one, over an observation interval lasting $T=300$ $T_{s}$;
these are denoted $RMSE_{L}($alg$)$ and $RMSE_{N}($alg$)$, respectively,
where `alg' denotes the algorithm these parameters refer to. Our assessment
of \emph{computational requirements} is based, instead, on assessing the
average \emph{execution time} required over the whole observation interval
(this quantity is denoted ET$($alg$)$ in the following). Moreover, the
following values have been selected for the parameters of the considered
SSM: $\rho =0.99$, $T_{s}=0.1$ s, $\sigma _{p}$ $=0.01$ m, $\sigma
_{e,p}=5\cdot 10^{-2}$ m, $\sigma _{e,v}=5\cdot 10^{-2}$ m/s, $a_{0}=1.5$ m/s%
$^{2}$, $d_{0}=0.5$ m, $\tilde{a}_{0}=0.05$ m/s$^{2}$ and $v_{0}=1$ m/s (the
initial position $\mathbf{p}_{0}\triangleq \lbrack p_{x,0},p_{y,0}]^{T}$ and
the initial velocity $\mathbf{v}_{0}\triangleq \lbrack v_{x,0},v_{y,0}]^{T}$
have been set to $[5$ m$,8$ m$]^{T}$ and $[4$ m/s$,$ $4$ m/s$]^{T}$,
respectively).

\begin{figure}[tbp]
\centering
\includegraphics[width=0.80%
\textwidth]{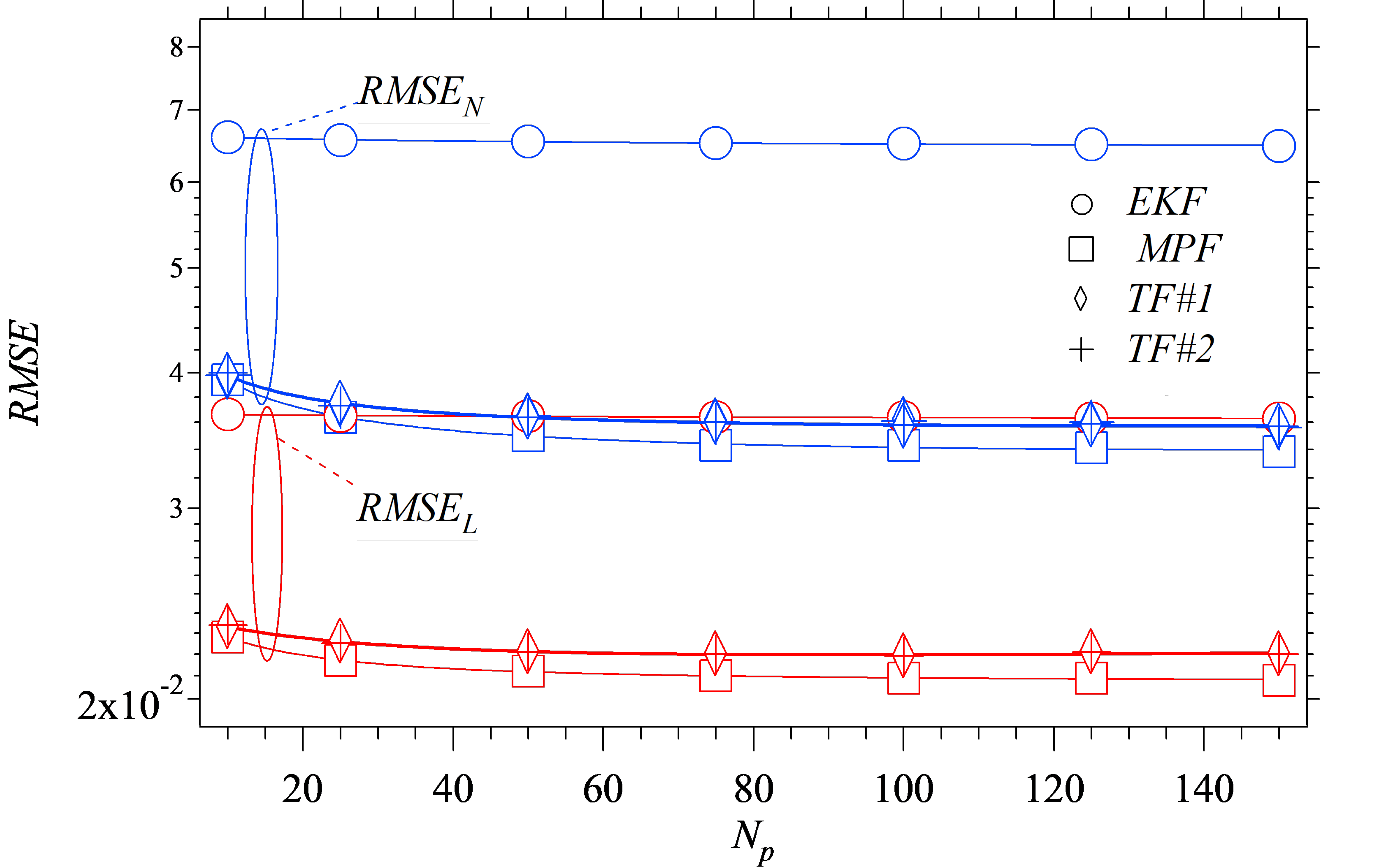}
\caption{RMSE performance versus $N_{p}$ for the linear component ($RMSE_{L}$%
) and the nonlinear component ($RMSE_{N}$) of system state; the CLG SSM
described by eqs. (\protect\ref{mod_1_v})-(\protect\ref{acc_SSM1}) and four
filtering techniques (EKF, MPF, TF\#1 and TF\#2) are considered.}
\label{Fig_1_sim}
\end{figure}

Some numerical results showing the dependence of $RMSE_{L}$ and $RMSE_{N}$
on the number of particles ($N_{p}$) for MPF, TF\#1 and TF\#2 are
illustrated in Fig. \ref{Fig_1_sim} (simulation results are indicated by
markers, whereas continuous lines are drawn to fit them, so facilitating the
interpretation of the available data); in this case $N_{it}=1$ has been
selected for both TF\#1 and TF\#2, and the range $[10,150]$ has been
considered for $N_{p}$. These results show that:

1) The value of $RMSE_{L}$ is significantly smaller than $RMSE_{N}$ for all
the algorithms; this is mainly due to the fact that the measurement vector $%
\mathbf{y}_{l}$ (\ref{mod_1_y}) provides richer information about $\mathbf{x}%
_{l}^{(L)}$ (i.e., $\mathbf{p}_{l}$) than about $\mathbf{x}_{l}^{(N)}$ ($%
\mathbf{v}_{l}$).

2) The EKF\ technique is appreciably outperformed by the other three
filtering algorithms in terms of both $RMSE_{L}$ and $RMSE_{N}$ for any
value of $N_{p}$; for instance, $RMSE_{L}($EKF$)$ ($RMSE_{N}($EKF$)$) is
about $1,65$\ ($1,80$)\ time larger than $RMSE_{L}($TF\#1$)$ ($RMSE_{N}($%
TF\#1$)$) for $N_{p}=100$.

3) Both TF\#1 and TF\#2 perform slightly worse than MPF for the same value
of $N_{p}$ (for instance, $RMSE_{L}($TF\#1$)$ and $RMSE_{N}($TF\#1$)$ are
about $5\%$ larger than the corresponding quantities evaluated for MPF);
moreover, there is no visible performance gap between TF\#1 and TF\#2, in
terms of both $RMSE_{L}$ and $RMSE_{N}$.

4) No real improvement in terms of $RMSE_{L}($alg$)$ and $RMSE_{N}($alg$)$
is found for $N_{p}\gtrsim 100$, if alg = MPF, TF\#1 or TF\#2

\begin{figure}[tbp]
\centering
\includegraphics[width=0.80%
\textwidth]{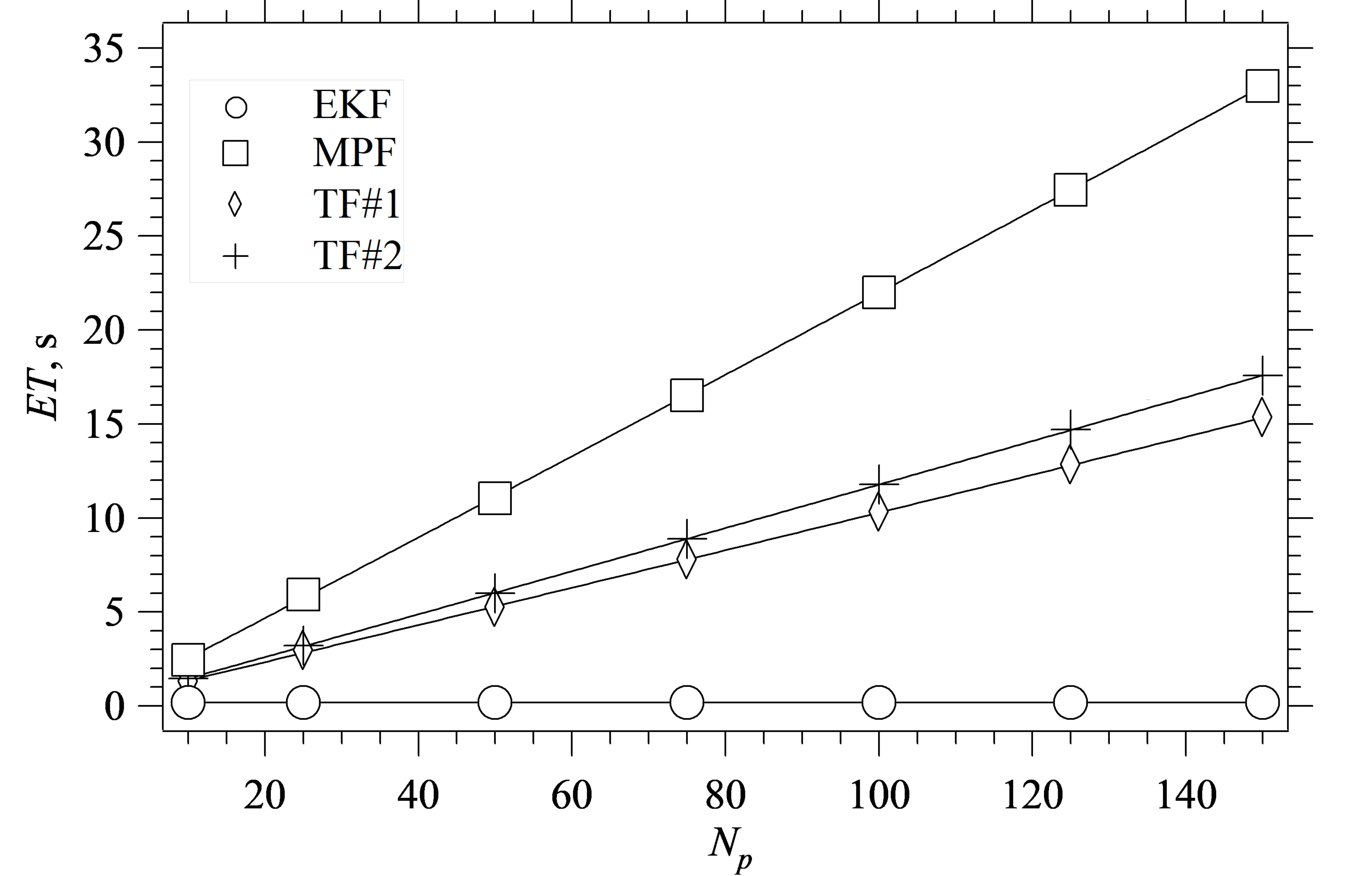}
\caption{ET versus $N_{p}$ for the EKF, MPF, TF\#1 and TF\#2; the CLG SSM
described by eqs. (\protect\ref{mod_1_v})-(\protect\ref{acc_SSM1}) is
considered.}
\label{Fig_2_sim}
\end{figure}

Despite their similar accuracies, MPF and TF algorithms require different
execution times; this is evidenced by the numerical results appearing in
Fig. \ref{Fig_2_sim} and showing the dependence of the ET parameter on $%
N_{p} $ for all the considered filtering algorithms. These results show that
TF\#1 and TF\#2 require an appreciably shorter execution time than MPF;
more precisely, the value of ET for TF1 (TF\#2) is approximately $0.61$ ($%
0.67$)$\ $times smaller than that required by MPF for the same value of $%
N_{p}$. Moreover, from Fig. \ref{Fig_1_sim}-\ref{Fig_2_sim} it is easily
inferred that, in the considered scanario, TF\#1 achieves a better RMSE -
ET tradeoff than both MPF and TF\#2.

Further simulation results (not shown here for space limitations) have also
evidenced that, in the considered scenario, no improvement in estimation
accuracy is obtained if $N_{it}>1$ is selected for TF\#1 and TF\#2.

\section{Conclusions\label{sec:conc}}

In this manuscript the concept of parallel concatenation of Bayesian filters
has been illustrated and a new graphical model has been developed for it.
This model can be exploited to develop a new family of filtering algorithms,
called \emph{turbo filters}. Two turbo filters have been derived for the
class of CLG SSMs and have been compared, in terms of both accuracy and
execution time, with EKF and MPF for a specific SSM. Simulation results
evidence that the devised TF schemes perform closely to MPF, but have
limited memory requirements and are appreciably faster.

\begin{appendices}
\section{\label{app:A}}

In this Appendix, the evaluation of the PM messages $\vec{m}_{pm,j}^{(k)}(%
\mathbf{x}_{l}^{(N)})$ (\ref{m_pm_x_N_l_j}) \ and $\vec{m}_{pm,j}^{(k)}(%
\mathbf{x}_{l}^{(L)})$ (\ref{message_pm_x_l_k}) is analysed in detail. 
\begin{figure}[tbp]
\centering
\includegraphics[width=0.80%
\textwidth]{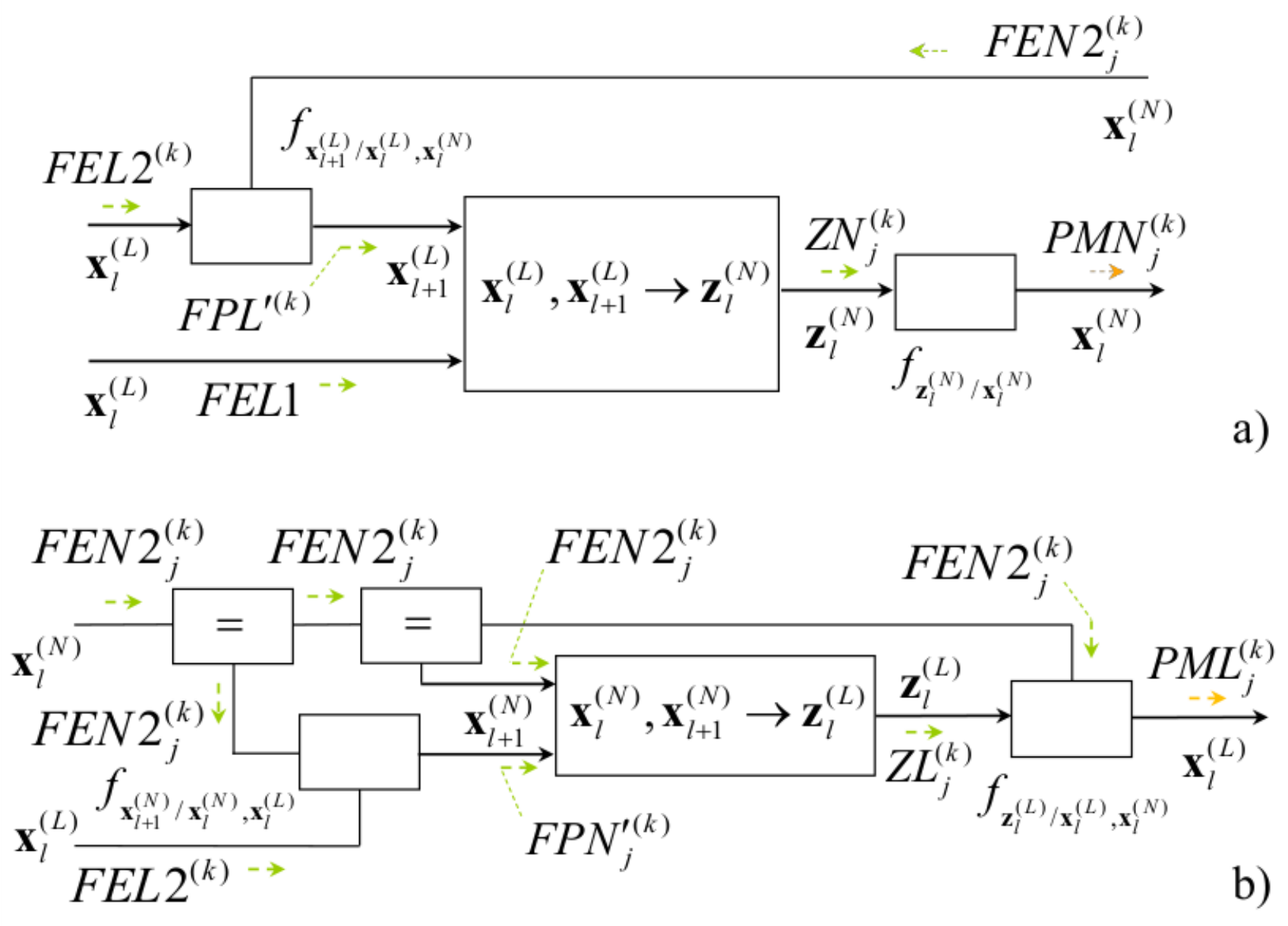}
\caption{Representation of the processing accomplished by a) the PMG$_{\text{%
EKF}}$ block and b) the PMG$_{\text{PF}}$ block (see Fig. \protect\ref{Fig_5}%
) as message passing over a FG.}
\label{Fig_7}
\end{figure}
The algorithm for computing $\vec{m}_{pm,j}^{(k)}(\mathbf{x}_{l}^{(N)})$ can
be represented as a message passing over the FG shown in Fig. \ref{Fig_7}%
-a). The expressions of the messages appearing in this graph can be derived
as follows. Given $\mathbf{x}_{l}^{(N)}=\mathbf{x}_{fp,l,j}^{(N)}[k-1]$
(conveyed by $\vec{m}_{fe2,j}^{(k)}(\mathbf{x}_{l}^{(N)})$ (\ref%
{m_fe_2_x_N_l_bis})) and $\vec{m}_{fe2}^{(k)}(\mathbf{x}_{l}^{(L)})$ (\ref%
{m_fe_L_EKF_2}), the message\footnote{%
The scale factor $w_{fe2,l,j}^{(k)}$ originating from $\vec{m}_{fe2,j}^{(k)}(%
\mathbf{x}_{l}^{(N)})$ (\ref{m_fe_2_x_N_l_bis}) can be ignored in the
following formula, since the resulting message is Gaussian \cite%
{Loeliger_2007}.} 
\begin{eqnarray}
\vec{m}_{fp,j}^{(k)}\left( \mathbf{x}_{l+1}^{(L)}\right) &=&\int \,f\left( 
\mathbf{x}_{l+1}^{(L)}\left\vert \mathbf{x}_{l}^{(L)},\mathbf{x}%
_{fp,l,j}^{(N)}[k-1]\right. \right) \,  \notag \\
&&\cdot \vec{m}_{fe2}^{(k)}(\mathbf{x}_{l}^{(L)})\,d\mathbf{x}_{l}^{(L)}
\label{eq:message_fp_L_l+1}
\end{eqnarray}%
providing a statistical representation of the prediction of $\mathbf{x}%
_{l+1}^{(L)}$ is computed first. Since $f(\mathbf{x}_{l+1}^{(L)}|\mathbf{x}%
_{l}^{(L)},\mathbf{x}_{fp,l,j}^{(N)}[k-1])=\mathcal{N}(\mathbf{x}%
_{l+1}^{(L)};\mathbf{f}_{l,j}^{(L)}[k]+\mathbf{A}_{l,j}^{(L)}[k]\,\mathbf{x}%
_{l}^{(L)}$, $\mathbf{C}_{w}^{(L)}\mathcal{)}$ (with $\mathbf{A}%
_{l,j}^{(L)}[k]\triangleq \mathbf{A}_{l}^{(L)}(\mathbf{x}%
_{fp,l,j}^{(N)}[k-1])$ and $\mathbf{f}_{l,j}^{(L)}[k]\triangleq \mathbf{f}%
_{l}^{(L)}(\mathbf{x}_{fp,l,j}^{(N)}[k-1])$), applying CR2 to the evaluation
of the integral in the RHS\ of (\ref{eq:message_fp_L_l+1}) produces 
\begin{equation}
\vec{m}_{fp,j}^{(k)}\left( \mathbf{x}_{l+1}^{(L)}\right) =\mathcal{N}(%
\mathbf{x}_{l+1}^{(L)};\mathbf{\tilde{\eta}}_{fp,l+1,j}^{(k)},\mathbf{\tilde{%
C}}_{fp,l+1,j}^{(k)}\mathcal{)}\,,  \label{eq:message_fp_L_l+1_bis}
\end{equation}%
where%
\begin{equation}
\mathbf{\tilde{\eta}}_{fp,l+1,j}^{(k)}\triangleq \mathbf{A}_{l,j}^{(L)}[k]%
\mathbf{\tilde{\eta}}_{fe2,l}^{(k)}+\mathbf{f}_{l,j}^{(L)}[k]
\label{eta_message_fp_L_l+1}
\end{equation}%
and%
\begin{equation}
\mathbf{\tilde{C}}_{fp,l+1,j}^{(k)}\triangleq \mathbf{C}_{w}^{(L)}+\mathbf{A}%
_{l,j}^{(L)}[k]\mathbf{\tilde{C}}_{fe2,l}^{(k)}\left( \mathbf{A}%
_{l,j}^{(L)}[k]\right) ^{T}.  \label{C_message_fp_L_l+1}
\end{equation}%
Then, the message $\vec{m}_{j}^{(k)}(\mathbf{z}_{l}^{(N)})$ is evaluated
(this message is denoted $ZN_{j}^{(k)}$ in Fig. \ref{Fig_7}-a)); this
expresses the pdf of $\mathbf{z}_{l}^{(N)}$ (\ref{z_N_l}) under the
assumptions that: a) $\mathbf{x}_{l}^{(N)}=\mathbf{x}_{fp,l,j}^{(N)}[k-1]$;
b) $\mathbf{x}_{l}^{(L)}$ and $\mathbf{x}_{l+1}^{(L)}$ are jointly Gaussian
vectors; c) the pdfs of $\mathbf{x}_{l}^{(L)}$ and $\mathbf{x}_{l+1}^{(L)}$
are expressed by $\vec{m}_{fe1}^{(k)}(\mathbf{x}_{l}^{(L)})$ (\ref%
{m_fe_L_EKF_1}) and $\vec{m}_{fp,j}^{(k)}(\mathbf{x}_{l+1}^{(L)})$ (\ref%
{eq:message_fp_L_l+1_bis}), respectively; d) the pdf of $\mathbf{x}%
_{l+1}^{(L)}$ conditioned on $\mathbf{x}_{l}^{(L)}$ and $\mathbf{x}%
_{l}^{(N)}=\mathbf{x}_{fp,l,j}^{(N)}[k-1]$ is $f(\mathbf{x}_{l+1}^{(L)}|%
\mathbf{x}_{l}^{(L)},\mathbf{x}_{fp,l,j}^{(N)}[k-1])=\mathcal{N}(\mathbf{x}%
_{l+1}^{(L)};\mathbf{f}_{l,j}^{(L)}[k]+\mathbf{A}_{l,j}^{(L)}[k]\mathbf{x}%
_{l}^{(L)},\mathbf{C}_{w}^{(L)})$ (see (\ref{eq:XL_update}) with $Z=L$).
Therefore, based on eq. (\ref{z_N_l}), the message $\vec{m}_{j}^{(k)}(%
\mathbf{z}_{l}^{(N)})$ can expressed as 
\begin{equation}
\vec{m}_{j}^{(k)}(\mathbf{z}_{l}^{(N)})=\mathcal{N}\left( \mathbf{z}%
_{l}^{(N)};\mathbf{\check{\eta}}_{z,l,j}^{(k)},\mathbf{\check{C}}%
_{z,l,j}^{(k)}\right) ,  \label{message_z_N}
\end{equation}%
where 
\begin{eqnarray}
\mathbf{\check{\eta}}_{z,l,j}^{(k)} &=&\mathbf{\tilde{\eta}}%
_{fp,l+1,j}^{(k)}-\mathbf{A}_{l,j}^{(L)}\mathbf{\tilde{\eta}}_{fe1,l}^{(k)} 
\notag \\
&=&\mathbf{A}_{l,j}^{(L)}\left[ \mathbf{\tilde{\eta}}_{fe2,l}^{(k)}-\mathbf{%
\tilde{\eta}}_{fe1,l}^{(k)}\right] +\mathbf{f}_{l,j}^{(L)}[k]
\label{eta_mess_z_N}
\end{eqnarray}%
and 
\begin{eqnarray}
\mathbf{\check{C}}_{z,l,j}^{(k)} &=&\mathbf{\tilde{C}}_{fp,l+1,j}^{(k)}-%
\mathbf{A}_{l,j}^{(L)}[k]\mathbf{\tilde{C}}_{fe1,l}^{(k)}\left( \mathbf{A}%
_{l,j}^{(L)}[k]\right) ^{T}  \notag \\
&=&\mathbf{C}_{w}^{(L)}+\mathbf{A}_{l,j}^{(L)}[k]\left[ \mathbf{\tilde{C}}%
_{fe2,l}^{(k)}-\mathbf{\tilde{C}}_{fe1,l}^{(k)}\right] \left( \mathbf{A}%
_{l,j}^{(L)}[k]\right) ^{T}.  \notag \\
&&  \label{C_mess_Z_N_bis}
\end{eqnarray}%
Finally, $\vec{m}_{j}^{(k)}(\mathbf{z}_{l}^{(N)})$ (\ref{message_z_N}) is
exploited to evaluate\footnote{%
Note that the following message represents the \emph{correlation} between
the pdf $\vec{m}_{j}(\mathbf{z}_{l}^{(N)})$ evaluated on the basis of the
definition (\ref{z_N_l}) and the pdf originating from the fact that this
quantity is expected to equal the random vector $\mathbf{f}_{l,j}^{(L)}+%
\mathbf{w}_{l}^{(L)}$ (see (\ref{z_N_l_bis})). For this reason, it expresses
the \emph{degree of similarity} between these two functions.} 
\begin{equation}
\vec{m}_{pm,j}^{(k)}\left( \mathbf{x}_{l}^{(N)}\right) =\int \vec{m}%
_{j}\left( \mathbf{z}_{l}^{(N)}\right) f\left( \mathbf{z}_{l}^{(N)}\left%
\vert \mathbf{x}_{fp,l,j}^{(N)}[k-1]\right. \right) d\mathbf{z}_{l}^{(N)}.
\label{m_pm_x_N_l_j_first}
\end{equation}%
Substituting (\ref{message_z_N}) and $f(\mathbf{z}_{l}^{(N)}|\mathbf{x}%
_{fp,l,j}^{(N)}[k-1])=\mathcal{N}(\mathbf{z}_{l}^{(N)};\mathbf{f}%
_{l,j}^{(L)}[k],\mathbf{C}_{w}^{(N)})$ (see (\ref{z_N_l_bis})) in the RHS of
the last expression and applying CR3 to the evaluation of the resulting
integral yields (\ref{m_pm_x_N_l_j}).

Similarly as $\vec{m}_{pm,j}^{(k)}(\mathbf{x}_{l}^{(N)})$, the algorithm for
computing the message $\vec{m}_{pm,j}^{(k)}(\mathbf{x}_{l}^{(L)})$ can be
represented as a message passing over a graphical model. Such a model is
shown in Fig. \ref{Fig_7}-b); moreover, the derivation of the messages
passed over it is sketched in the following. Given $\mathbf{x}_{l}^{(N)}=\mathbf{x}%
_{fp,l,j}^{(N)}[k]$ (conveyed by the message $\vec{m}_{fe2,j}^{(k)}(\mathbf{x%
}_{l}^{(N)})$ (\ref{m_fe_2_x_N_l_tris})) and $\vec{m}_{fe2}^{(k)}(\mathbf{x}%
_{l}^{(L)})$ (\ref{m_fe_L_EKF_2}), the message%
\begin{eqnarray}
\vec{m}_{fp,j}^{(k)}\left( \mathbf{x}_{l+1}^{(N)}\right)  &=&\int \int
\,f\left( \mathbf{x}_{l+1}^{(N)}\left\vert \mathbf{x}_{l}^{(L)},\mathbf{x}%
_{fp,l,j}^{(N)}[k]\right. \right)   \notag \\
&&\cdot \vec{m}_{fe2}^{(k)}\left( \mathbf{x}_{l}^{(L)}\right) \,d\mathbf{x}%
_{l}^{(L)},  \label{eq:double_integral_1}
\end{eqnarray}%
representing a forward prediction of $\mathbf{x}_{l+1}^{(N)}$, is evaluated
first.\ Applying CR2 to the evaluation of the last integral (note that $f(%
\mathbf{x}_{l+1}^{(N)}|\mathbf{x}_{fp,l,j}^{(N)}[k],\mathbf{x}_{l}^{(L)})=%
\mathcal{N}(\mathbf{x}_{l+1}^{(N)};\mathbf{A}_{l,j}^{(N)}[k]\mathbf{x}%
_{l}^{(L)}+\mathbf{f}_{l,j}^{(N)}[k],\mathbf{C}_{w}^{(N)})$, with $\mathbf{A}%
_{l,j}^{(N)}[k]\triangleq \mathbf{A}_{l}^{(N)}(\mathbf{x}_{fp,l,j}^{(N)}[k])$
and $\mathbf{f}_{l,j}^{(N)}[k]\triangleq \mathbf{f}_{l}^{(N)}(\mathbf{x}%
_{fp,l,j}^{(N)}[k])$, and that $\vec{m}_{fe2}^{(k)}(\mathbf{x}_{l}^{(L)})$ (\ref%
{m_fe_L_EKF_2}) is Gaussian) yields%
\begin{equation}
\vec{m}_{fp,j}^{(k)}\left( \mathbf{x}_{l+1}^{(N)}\right) =\mathcal{N}\left( 
\mathbf{x}_{l+1}^{(N)};\mathbf{\check{\eta}}_{fp,l,j}^{(k)},\mathbf{\check{C}%
}_{fp,l,j}^{(k)}\right) ,  \label{eq:m_fp_N_l+1_bis}
\end{equation}%
where%
\begin{equation}
\mathbf{\check{\eta}}_{fp,l,j}^{(k)}\triangleq \mathbf{A}_{l,j}^{(N)}[k]%
\mathbf{\tilde{\eta}}_{fe2,l}^{(k)}+\mathbf{f}_{l,j}^{(N)}[k]
\label{eq:eta_fp_N_l+1}
\end{equation}%
and%
\begin{equation}
\mathbf{\check{C}}_{fp,l,j}^{(k)}\triangleq \mathbf{C}_{w}^{(N)}+\mathbf{A}%
_{l,j}^{(N)}[k]\mathbf{\tilde{C}}_{fe2,l}^{(k)}\left( \mathbf{A}%
_{l,j}^{(N)}[k]\right) ^{T}.  \label{eq:C_fp_N_l+1}
\end{equation}%
Then, the message $\vec{m}_{fp,j}^{(k)}(\mathbf{x}_{l+1}^{(N)})$ (\ref%
{eq:m_fp_N_l+1_bis}) is replaced by its \emph{particle-based representation}%
; this result is achieved sampling the Gaussian function $\mathcal{N}(%
\mathbf{x}_{l+1}^{(N)};\mathbf{\check{\eta}}_{fp,l,j}^{(k)},\mathbf{\check{C}%
}_{fp,l,j}^{(k)})$ (see (\ref{eq:m_fp_N_l+1_bis})), that is drawing the
sample $\mathbf{x}_{fp,l+1,j}^{(N)}[k]$ from it and b) assigning the weight $%
1/N_{p}$ to this sample. The value of the PM $\mathbf{z}_{l}^{(L)}$ (\ref%
{eq:z_L_l}) associated with the couple $(\mathbf{x}_{l}^{(N)},\mathbf{x}%
_{l+1}^{(N)})=(\mathbf{x}_{fp,l,j}^{(N)}[k]$ , $\mathbf{x}%
_{fp,l+1,j}^{(N)}[k])$ is%
\begin{equation}
\mathbf{z}_{l,j}^{(L)}\left[ k\right] \triangleq \mathbf{x}%
_{fp,l+1,j}^{(N)}[k]-\mathbf{f}_{l,j}^{(N)}[k]  \label{eq:z_L_evaluated}
\end{equation}%
and is conveyed by the message (denoted $ZL_{j}^{(k)}$ in Fig. \ref{Fig_7}-b)%
\begin{equation}
\vec{m}_{j}^{(k)}\left( \mathbf{z}_{l}^{(L)}\right) =\delta \left( \mathbf{z}%
_{l}^{(L)}-\mathbf{z}_{l,j}^{(L)}\left[ k\right] \right) .
\label{m_j_z_L_bis}
\end{equation}%
Then, the message $\vec{m}_{pm,j}^{(k)}(\mathbf{x}_{l}^{(L)})$ is evaluated
as (see Fig. \ref{Fig_7}-b)) 
\begin{equation}
\vec{m}_{pm,j}^{(k)}\left( \mathbf{x}_{l}^{(L)}\right) =\int \vec{m}%
_{j}^{(k)}\left( \mathbf{z}_{l}^{(L)}\right) \,f\left( \mathbf{z}%
_{l}^{(L)}\left\vert \mathbf{x}_{l}^{(L)},\mathbf{x}_{l}^{(N)}\right.
\right) d\mathbf{z}_{l}^{(L)}.  \label{eq:message_pm_L_j}
\end{equation}%
Substituting (\ref{m_j_z_L_bis}) and $f(\mathbf{z}_{l}^{(L)}|\mathbf{x}%
_{l}^{(L)},\mathbf{x}_{l}^{(N)})=\mathcal{N(}\mathbf{z}_{l}^{(L)}$; $\mathbf{%
A}_{l,j}^{(N)}[k]\mathbf{x}_{l}^{(L)},\mathbf{C}_{w}^{(N)})$ (see (\ref%
{eq:z_L_l_bis})) in the RHS of (\ref{eq:message_pm_L_j}) yields the message $%
\vec{m}_{pm,j}^{(k)}(\mathbf{x}_{l}^{(L)})=\mathcal{\mathcal{N(}}\mathbf{z}%
_{l,j}^{(L)}\left[ k\right] ;\mathbf{A}_{l,j}^{(N)}\left[ k\right] \mathbf{x}%
_{l}^{(L)},\mathbf{C}_{w}^{(N)})$, that can be easily put in the equivalent
Gaussian form (\ref{eq:message_pm_L_j_tris}).
\end{appendices}

\end{document}